\documentclass[acmtog, noacm]{acmart}

\renewcommand\footnotetextcopyrightpermission[1]{} 
\usepackage{booktabs} 
\usepackage[ruled]{algorithm2e} 

\SetAlFnt{\small}
\SetAlCapFnt{\small}
\SetAlCapNameFnt{\small}
\SetAlCapHSkip{0pt}
\usepackage{multirow}
\usepackage{relsize}
\usepackage{colortbl}
\usepackage{graphicx}
\usepackage{subfigure}

\newcommand{\hl}[1]{{#1}}

\begin{document}
\title{Physical Non-inertial Poser (PNP): Modeling Non-inertial Effects in Sparse-inertial Human Motion Capture}

\author{Xinyu Yi}
\orcid{0000-0003-3504-3222}
\affiliation{%
  \institution{School of Software and BNRist, Tsinghua University}
  \city{Beijing}
  \country{China}}
\email{yixy20@mails.tsinghua.edu.cn}

\author{Yuxiao Zhou}
\orcid{0009-0005-6189-2326}
\affiliation{%
  \institution{ETH Zurich}
  \city{Zurich}
  \country{Switzerland}}
\email{yuxiao.zhou@outlook.com}

\author{Feng Xu}
\orcid{0000-0002-0953-1057}
\affiliation{%
  \institution{School of Software and BNRist, Tsinghua University}
  \city{Beijing}
  \country{China}}
\email{xufeng2003@gmail.com}

\begin{abstract}
    Existing inertial motion capture techniques use the human root coordinate frame to estimate local poses and treat it as an inertial frame by default.
We argue that when the root has linear acceleration or rotation, the root frame should be considered non-inertial theoretically.
In this paper, we model the fictitious forces that are non-neglectable in a non-inertial frame by an auto-regressive estimator delicately designed following physics. 
With the fictitious forces, the force-related IMU measurement (accelerations) can be correctly compensated in the non-inertial frame and thus Newton's laws of motion are satisfied.
In this case, the relationship between the accelerations and body motions is deterministic and learnable, and we train a neural network to model it for better motion capture.
Furthermore, to train the neural network with synthetic data, we develop an IMU synthesis by simulation strategy to better model the noise model of IMU hardware and allow parameter tuning to fit different hardware.
This strategy not only establishes the network training with synthetic data but also enables calibration error modeling to handle bad motion capture calibration, increasing the robustness of the system.
Code is available at \hl{\url{https://xinyu-yi.github.io/PNP/}}.

\end{abstract}

\begin{teaserfigure}
  \includegraphics[width=\textwidth]{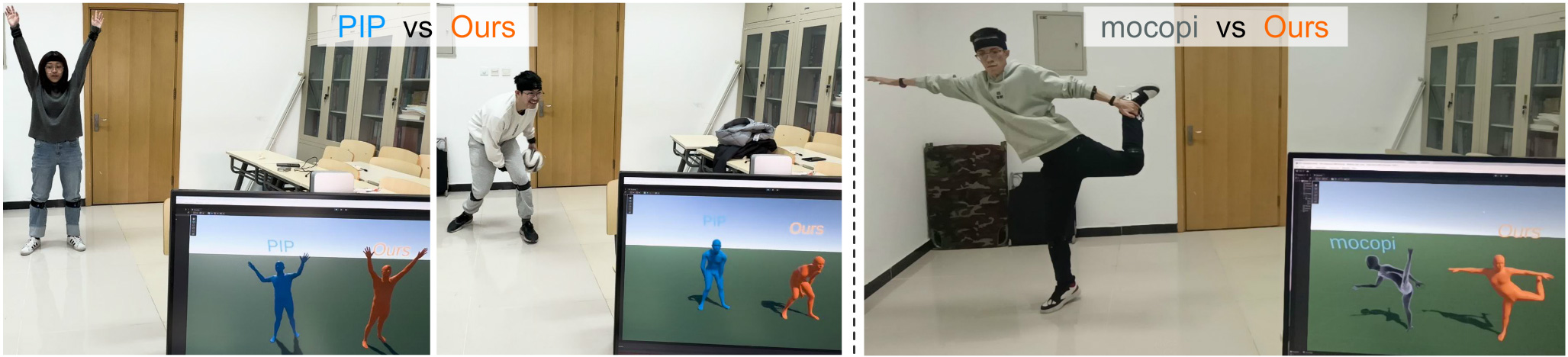}
  \caption{Live comparisons between our method and the state-of-the-art techniques in both academia (PIP, left) and industry (mocopi from Sony, right) in motion capture from six wearable Inertial Measurement Units (IMUs). Our approach outperforms previous works on ambiguous (left) and complex poses (right).}
  \label{fig:teaser}
\end{teaserfigure}

\maketitle
\fancyfoot{}
\thispagestyle{empty}

\section{Introduction}
Human motion capture is a long-standing research interest.
The evolution of MoCap (motion capture) technology keeps presenting new possibilities in various fields such as film, animation, gaming, sports, and rehabilitation.
\par
For decades, MoCap with \textit{dense} optical markers~\cite{Vicon} has been the golden standard in terms of accuracy, but it suffers from expensive system setup and the requirement of a dedicated studio due to its fragility to the surroundings.
With the advancement of inertial sensing technology, MoCap with inertial measurement units (IMUs) commenced to emerge.
Compared to optical solutions, IMU-based systems are expense-friendly and immune to occlusion, visibility, or lighting.
These advantages render them suitable in unconstrained recording environments.
While MoCap with \textit{dense} IMUs has been industrialized~\cite{Xsens,Noitom}, pioneering works with \textit{sparse} IMUs have also achieved promising MoCap quality~\cite{SIP,DIP,PIP,TIP,TransPose}.
By reducing the number of IMUs from 17 to only 6, the system becomes available for end users, and user comfort during capture is significantly enhanced.
Since IMUs are already integrated into common smart devices such as phones, watches, and earbuds, MoCap with sparse IMUs is a steady step towards everyday motion understanding~\cite{imuposer,lee2024mocap}.
%

%
%
%
\par
However, we identify a physical flaw in the previous IMU-based methods like~\cite{DIP,PIP,TIP,TransPose}, which leads to the failure of reconstructing some challenging motions.
We illustrate this flaw as follows.
For easy modeling and calculation, the task of MoCap is typically decoupled into two components: root-relative pose estimation and global movement integration.
To estimate the local pose, previous methods first transform the IMU measurements from the world frame to the human's root frame according to the rotation matrix read from the root IMU.
Since the world frame is an inertial frame, such transformation is physically correct only if the root frame is also inertial.
However, during human movements, the root frame is usually non-inertial as the root joint keeps experiencing acceleration and rotation.
In such cases, projecting IMU measurements from the inertial world frame to the non-inertial root frame requires introducing \textit{fictitious forces}, such as centrifugal force and Coriolis force~\cite{dynacc}, to compensate for the raw acceleration measurements from the IMU sensors.
In previous works, the neglect of these forces during the transformation results in "wrong" acceleration measurements that do not match the actual local movements from the perspective of the non-inertial root frame.
As a result, previous models could not learn the correlation between acceleration and human movement well, and in fact discarded most acceleration information, as reported in previous works~\cite{DIP,TransPose}.
While these works can still estimate many local poses only using the orientation measurements from the IMU, they face difficulties in handling specific motions where the orientation readings remain almost unchanged and the dynamics can only be observed from accelerations, \textit{e.g.}, raising hands or legs without rotating them.
\par
In this paper, we explicitly model the fictitious forces induced by the world-to-local transform to correct the IMU acceleration readings for local pose estimation.
Note that the fictitious forces are not real forces, but rather serve to mathematically describe the motion of objects in non-inertial frames~\cite{dynacc}. 
Our method rigorously involves the calculation of fictitious forces within the non-inertial root coordinate system of the human body.
Building upon this formulation, we designed a novel auto-regressive neural estimator that additionally takes the variables associated with the fictitious forces from the mathematical formulation as inputs to model the compensation forces effectively. 
Thanks to supervised learning with the \textit{correct} acceleration in the non-inertial root frame, the neural estimator successfully integrates the physical and data-driven aspects, enabling it to capture the dynamics of the fictitious forces. 
Subsequently, the force is utilized to modulate the raw acceleration measurement from the IMUs. 
This innovative approach enables our model to learn and utilize acceleration for motion estimation, thereby achieving tracking capabilities beyond the scope of previous works.
\par
While modeling the fictitious forces should improve the capture performance theoretically, it implicitly requires a large amount of paired real acceleration signals and ground truth poses for training, which are difficult to obtain for the entire research community.
%
%
At the moment, most previous methods first trained their models on large motion capture datasets with synthetic sensor measurements and then fine-tuned the models on smaller real datasets~\cite{DIP,TransPose,PIP}.
These works synthesize the sensor input by directly using the ground truth \textit{motion measurements}, \textit{e.g.}, rotation matrix from the known human pose.
This approach is suboptimal because modeling the real-world noise for rotations is extremely hard, and these noise patterns will differ with different IMU manufacturers due to different firmware algorithms.
As a result, the domain gap still deeply hurts their motion capture quality in the real world.
To this end, we propose a new method that directly synthesizes the \textit{raw IMU signals} (\textit{i.e.}, linear accelerations, angular velocities, and magnetic field measurements), and employ our sensor fusion algorithm to obtain the final motion measurements.
By leveraging the noise model at the hardware level~\cite{orbslam2,vinsmono}, we can use an examined set of noise parameters to generate signals more similar to real use cases.
This is particularly useful when the users already know the noise parameters of their sensors.
%
%
Finally, we model calibration errors introduced in the T-pose calibration in inertial motion capture.
Unlike previous approaches that assume perfect IMU calibration, we incorporate these errors in our synthetic IMU measurements and thus significantly reduce the error during real-world usage.
\par
%
%
%
%
%
%
%
In conclusion, our main contributions include:
\begin{itemize}
    \item Physical Non-inertial Poser (PNP), which enhances real-time human motion estimation from sparse inertial measurement units (IMUs), especially for acceleration-dominated motions like raising hands or legs.
    \item Fictitious force modeling is accomplished through a neural auto-regressive estimator that learns the physically correct fictitious forces arising from modeling the non-inertial human root's coordinate frame.
    \item IMU measurement synthesis by simulation which generates more realistic IMU signals from motion capture data while considering sensor noise and calibration errors.
\end{itemize}
\par
%
%
%
\section{Related Work}
\subsection{Motion Capture from Body-worn Sensors}
In this section we introduce previous MoCap works using body-mounted sensors.
These works can be categorized into two groups by the sensor type: 6DoF trackers that provide global locations using external stations, and IMUs without any positional information.
\par
6DoF trackers are commonly used in virtual reality systems, such as head-mounted displays and hand-held controllers.
By placing base stations or infrared cameras in the captured volume, these trackers can provide positional information in addition to orientations.
Numerous studies have explored human full-body tracking using sparse 6DoF trackers, with configurations ranging from 6 trackers placed on the end-effects and the root~\cite{sparseposer}, to 4 trackers on the upper body~\cite{lobstr}, and even just 3 trackers for the hands and the head (typically a head-mounted device (HMD) plus two hand-held controllers)~\cite{dittadi2021full,avatarposer,ahuja2021coolmoves,aliakbarian2022flag,agrol,neural3points,Aliakbarian_2023_ICCV,Zheng_2023_ICCV,jiang2023egoposer,shin2023utilizing,castillo2023bodiffusion}.
Some research incorporates physics-based simulation to ensure the physical correctness of the captured motion~\cite{questsim,questenvsim,neural3points}.
While these techniques yield promising results, they are inherently limited by the requirement of external devices for positioning.
The calibration of external devices is inconvenient, and the capture environment is constrained.
As a result, these methods are not applicable for long-range captures in an open world, such as outdoor movements.
\par
While the use of 6DoF trackers relies on external stations or cameras, using IMUs allows free-range movements. 
IMUs can measure orientation and acceleration without requiring external devices, but they do not provide localization.
Due to the lack of position information, thoroughly exploiting orientation and acceleration is vital for IMU-based MoCap methods.
Current commercial solutions~\cite{Xsens,Noitom} integrate dense IMUs into a tight suit, which is uncomfortable, intrusive, and hinders the user's movements.
Recent works focus on MoCap with fewer IMUs.
To address the pose ambiguity due to the reduction of IMUs, some works~\cite{robustcap,hybridcap,Pons2010,EgoLocate} assist the capture with additional external or egocentric hardware.
In the following, we only discuss pure IMU methods. 
SIP~\cite{SIP} first proposed reconstructing human pose from solely 6 IMUs through offline optimization.
DIP~\cite{DIP} further extended the technique to real-time performance by using deep recurrent neural networks.
While the work only provides local poses, TransPose~\cite{TransPose} estimates global translation by constraining foot-ground contacts.
The state-of-the-art works PIP~\cite{PIP} and TIP~\cite{TIP} refine TransPose from two aspects: 
PIP incorporates physics-based optimization to ensure the captured motion being physically plausible, while TIP adopts the transformer architecture~\cite{vaswani2017attention} to improve tracking accuracy.
While these works demonstrate promising results, their pose estimation components mostly rely on orientation signals, but cannot utilize acceleration readings correctly, as reported in DIP and TransPose.
As a result, they fail to track acceleration-dependent motions like raising hands, which are not identifiable solely by the orientations from sparse IMUs.
In this work, we introduce a new physical model that utilizes acceleration in a physically correct way.
\subsection{Virtual IMU Measurement Generation}
While MoCap datasets are abundant, most of them only contain clean and smooth human motion without original raw IMU signals.
In previous works~\cite{TransPose,PIP,DIP}, the synthetic IMU measurements are calculated by placing virtual IMUs on corresponding body parts using the MoCap sequences.
However, even fine-tuned on small real datasets, these models still hurt from the domain gap between synthetic ideal IMUs and real-world noisy IMUs.
The synthesis of virtual IMU measurements is crucial for this task. 
In the pioneering work of~\cite{DIP}, orientation measurements are synthesized from bone orientations, while acceleration measurements are derived by applying finite differences to the IMU positions.
However, subsequent works~\cite{TransPose,PIP,TIP} found that the acceleration measurements generated by finite differences exhibit a significant disparity between synthetic and real signals.
To address this issue, \cite{TransPose} and \cite{PIP} increased the step length in the finite difference algorithm to smooth the values, while \cite{TIP} suggested using an averaging filter with a sliding window for both synthetic and real data to minimize the gap.
However, they both inadvertently discard the high-frequency details in accelerations, which directly relate to fast and sudden movements.
On the other hand, the discrepancy between synthetic and real IMU measurements still persists as the sensor noises and calibration errors are not considered.
Virtual IMU measurement generation has also been explored in other tasks such as human action recognition from  videos~\cite{rey2019let,imutube} or motion sequences~\cite{xiao2021deep,takeda2018multi}.
Inspired by the works of \cite{imutube,imusim}, in this work, we directly synthesize the raw IMU signals to simulate the sensor noise in real world, and introduce a calibration error model to further improve the results.
\begin{figure}
    \includegraphics[width=\linewidth]{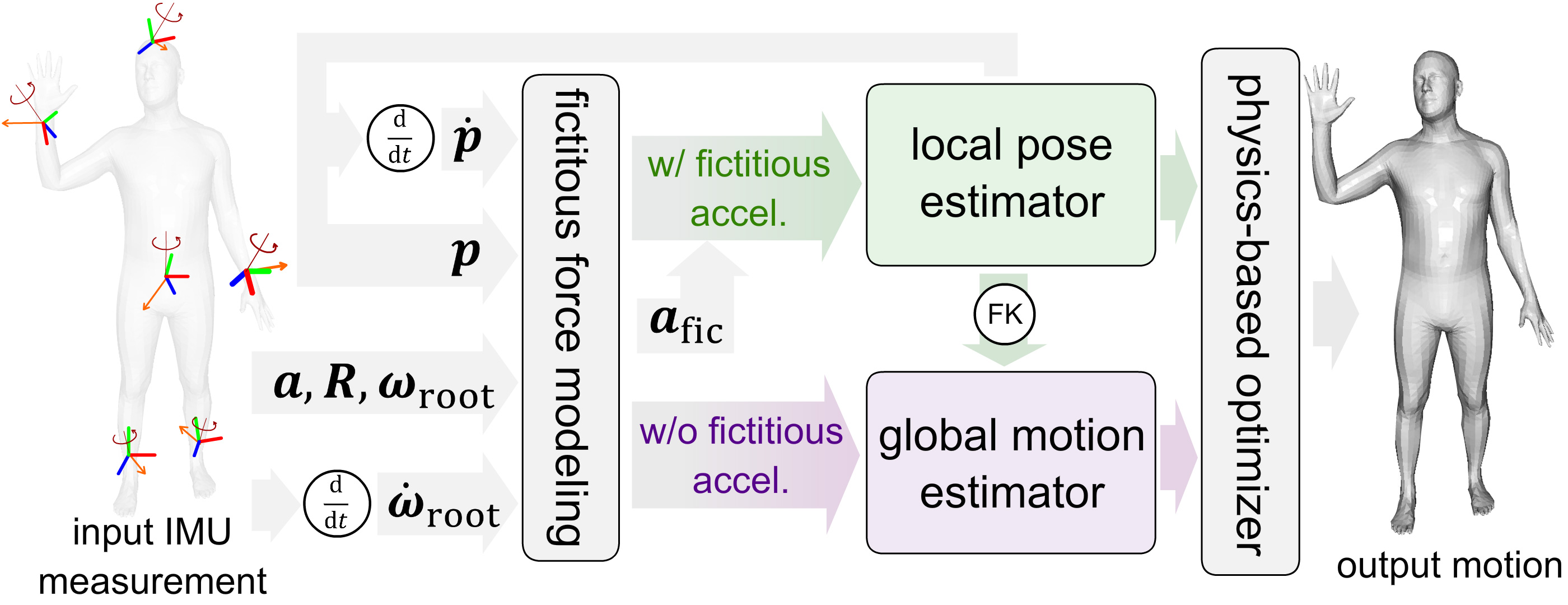}
    \caption{
        Overview of our motion capture method. 
        We first estimate the fictitious force in the non-inertial root frame (grey, left). 
        Then, we transform the IMU measurements from the world frame to the root frame to estimate the local pose, while accounting for the fictitious force (green).
        Next, we estimate the human's global motion from the pose and the IMUs (purple).
        Finally, we employ a physics-based optimizer to refine the human motion (grey, right).
    }
    \label{fig:methoda}
\end{figure}
\section{Method}
Our goal is to estimate real-time human motion from 6 Inertial Measurement Units (IMUs) placed on the leaf joints (forearms, lower legs, head) and the root joint (pelvis).
%
%
The inputs to our system are IMU measurements including accelerations, angular velocities, and orientations.
The outputs of our system are the local pose and global movements.
In the following, we first introduce our novel system (Sec.~\ref{sec:3.1}) that incorporates fictitious forces in the non-inertial coordinate frame, then explain how we synthesize realistic IMU data (Sec.~\ref{sec:3.2}) to fuel our models.
See Fig.~\ref{fig:methoda} and \ref{fig:methodb} for an overview.
\subsection{Framework Design}\label{sec:3.1}
Estimating human motion from IMU measurements is challenging due to the inherent coupling between local motions and global movements in sensor signals.
To this end, our method disentangles human local pose and global translation estimation into separate tasks, as in previous works~\cite{TransPose,PIP}. 
To estimate local pose, we transform IMU measurements from the global frame to the root-relative frame according to the orientation readings from the sensor on the root joint.
While the global frame is inertial, the root frame is often non-inertial due to body movements.
When projecting IMU measurements, especially acceleration signals, from an inertial frame to a non-inertial frame, fictitious forces need to be introduced as compensations.
While prior works neglected them, in Sec.~\ref{sec:3.1.1} we explain how these fictitious forces are modeled in our method.
Then, in Sec.~\ref{sec:3.1.2} we introduce how our method estimates local pose and global motion with the corrected acceleration.
%
%
%
\begin{figure}
    \centering
    \includegraphics[width=\linewidth]{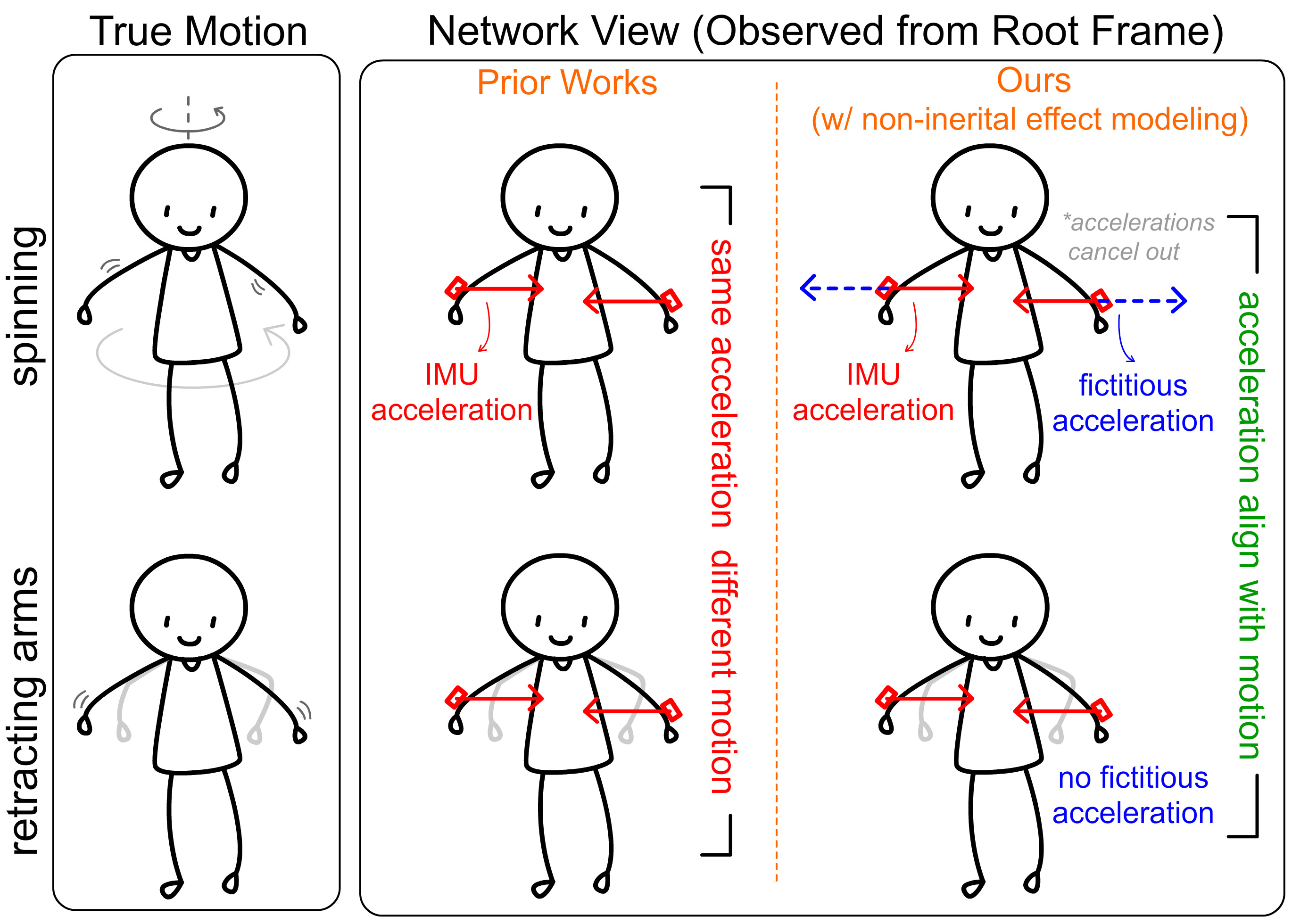}
    \caption{Illustration of the necessity of non-inertial effects modeling in human local pose estimation. Without fictitious accelerations, two different motions (left) own the same acceleration observations in the root frame (middle) while the accelerations are correctly observed with the consideration of the fictitious accelerations (right).}
    \label{fig:illustration}
\end{figure}
\subsubsection{Fictitious Force Modeling}\label{sec:3.1.1}
%
%
%
In the non-inertial human's root coordinate system, fictitious forces will affect the inertia measurements and must be accounted for when estimating the root-relative human pose (Fig.~\ref{fig:illustration}). 
Formally, we denote the acceleration and the angular velocity of the root joint as $\boldsymbol{a}_{RR}$\footnote{Here the notation $x_{AB}$ denotes the quantity $x$ of the origin of the coordinate frame $B$ expressed in the coordinate frame $A$. For instance, we use $\boldsymbol{p}_{AB}$ to represent the position of the origin of the coordinate frame B expressed in frame A, and $\boldsymbol{v}_{AB}$ to represent the velocity of the origin of the coordinate frame B expressed in frame A. However, when the quantity refers to the rotation $R$, $\boldsymbol{R}_{AB}$ denotes the rotation of the coordinate frame B with respect to the frame A. We consistently apply this subscript rule throughout the paper. Here, $\boldsymbol{a}_{RR}$ denotes the acceleration of the origin of the root frame (as $R$ stands for the root frame) expressed in the root frame.} and $\boldsymbol{\omega}_{RR}$ respectively.
%
Any leaf joint $L$ with position $\boldsymbol{p}_{RL}$ and velocity $\dot{\boldsymbol{p}}_{RL}$ in the human's root frame is subject to a fictitious force $\boldsymbol{f}_{\mathrm{fic}}$:
\begin{equation}\label{eq:fictitious-force}
    \boldsymbol{f}_{\mathrm{fic}} = -m(\boldsymbol{a}_{RR} + [\boldsymbol{\omega}_{RR}]_\times^2\boldsymbol{p}_{RL} + \\
    2[\boldsymbol{\omega}_{RR}]_\times\dot{\boldsymbol{p}}_{RL} + [\dot{\boldsymbol{\omega}}_{RR}]_\times\boldsymbol{p}_{RL}),
\end{equation}
where $m$ is the mass and $[\cdot]_\times$ is the skew-symmetric matrix for the vector cross product. The fictitious force consists of four terms, namely the linear inertial force, centrifugal force, Coriolis force, and Euler force~\cite{dynacc}, corresponding to the terms in Eq.~\ref{eq:fictitious-force}. 
\hl{Readers can refer to the supplemental paper for the derivation.}
This equation indicates that when observed from the root coordinate frame, the acceleration of a leaf joint does not equal the raw IMU measurements evaluated in the inertial world frame.
Instead, it undergoes an additional acceleration caused by the fictitious force $\boldsymbol{f}_{\mathrm{fic}}$.
We refer to this additional acceleration as \textit{fictitious acceleration}.
\par
In the following, we explain how to compute the fictitious acceleration for an arbitrary leaf joint.
As shown in Eq.~\ref{eq:fictitious-force}, the fictitious acceleration depends on the root joint acceleration $\boldsymbol{a}_{RR}$ and angular velocity $\boldsymbol{\omega}_{RR}$, as well as leaf joint position $\boldsymbol{p}_{RL}$ and velocity $\dot{\boldsymbol{p}}_{RL}$.
While the root joint terms can be read from the corresponding IMU, the leaf joint terms are not directly measured by the sensors and have to be computed using the estimated body pose.
%
%
Instead of calculating the fictitious acceleration manually from Eq.~\ref{eq:fictitious-force}, we employ an autoregressive neural network to estimate the acceleration.
The input to the network comprises root joint dynamics and leaf joint dynamics.
Root joint dynamics are acceleration $\boldsymbol{a}_{RR}$, angular velocity $\boldsymbol{\omega}_{RR}$, and angular acceleration $\dot{\boldsymbol{\omega}}_{RR}$.
Leaf joint dynamics include position $\boldsymbol{p}_{RL}$, velocity $\dot{\boldsymbol{p}}_{RL}$, acceleration $\boldsymbol{a}_{RL}$, and orientation $\boldsymbol{R}_{RL}$.
Note that the leaf joint positions and velocities are estimations from the \textit{previous frame}, and the network implicitly learns to wrap these values to the current frame.
All these values are concatenated into one vector as the network input, and the output is the fictitious acceleration $\boldsymbol{a}_{\mathrm{fic}}=\boldsymbol{f}_{\mathrm{fic}}/m$ for the corresponding leaf joint.
In practice, we use one fused network for all 5 leaf joints, \textit{i.e.,} the inputs are the dynamics of the root joint and all leaf joints, while the outputs are the fictitious accelerations for them.
%
%
We implement the estimator as a fully connected neural network trained with L2 loss.
For supervision, we synthesize the ground-truth fictitious acceleration from the motion capture data.
\par
\hl{We compute the fictitious acceleration using a network rather than Eq.~\ref{eq:fictitious-force} is based on three considerations. First, we need the network to forecast the unknown leaf joint positions and velocities from the values of the previous frame. Second, the root joint's angular velocity is prone to measurement noise, which can be mitigated by leveraging a network. Third, Eq.~\ref{eq:fictitious-force} individually computes the fictitious acceleration of each leaf joint, whereas a network can estimate all leaf joints altogether, incorporating human motion priors.}
%
%
%
\subsubsection{Local Pose and Global Motion Estimation}\label{sec:3.1.2}
With the estimated fictitious acceleration $\boldsymbol{a}_{\mathrm{fic}}$, now we can correctly transform the input IMU measurements to the human's root frame for local pose estimation.
Formally, the input is the concatenated vector of $\{\boldsymbol{R}_{RL}, \boldsymbol{a}_{RL} + \boldsymbol{a}_{\mathrm{fic}}\}$ containing leaf joint rotation matrices and accelerations in the non-inertial root frame.
Following~\cite{TransPose,PIP}, we first estimate leaf joint positions $\boldsymbol{p}_{RL}$, then all joint positions, and finally all joint rotations.
This is achieved through 3 Long Short-Term Memory (LSTM) recurrent networks~\cite{LSTM} with jump connections on the IMU input~\cite{TransPose,PIP}.
The intermediate estimation of the leaf joint position $\boldsymbol{p}_{RL}$, together with its time derivative $\dot{\boldsymbol{p}}_{RL}$ computed by finite difference, are fed back to the fictitious force estimator for the next frame's use.
\par
For global translation estimation, we follow~\cite{PIP} to regress joint velocities and foot-ground contact probabilities.
%
Note that different from the local pose estimation, fictitious forces are \textit{not} needed in the global motion estimation as we regress the global movement with respect to the static world frame, which is an inertial frame of reference.
We finally perform a physics-based motion optimization to ensure the physical correctness of the captured motion as in~\cite{PIP}.
%
%
%
%
%
%
%
%
%
%
%
%
%
%
%
%
%
%
\begin{figure}
    \includegraphics[width=\linewidth]{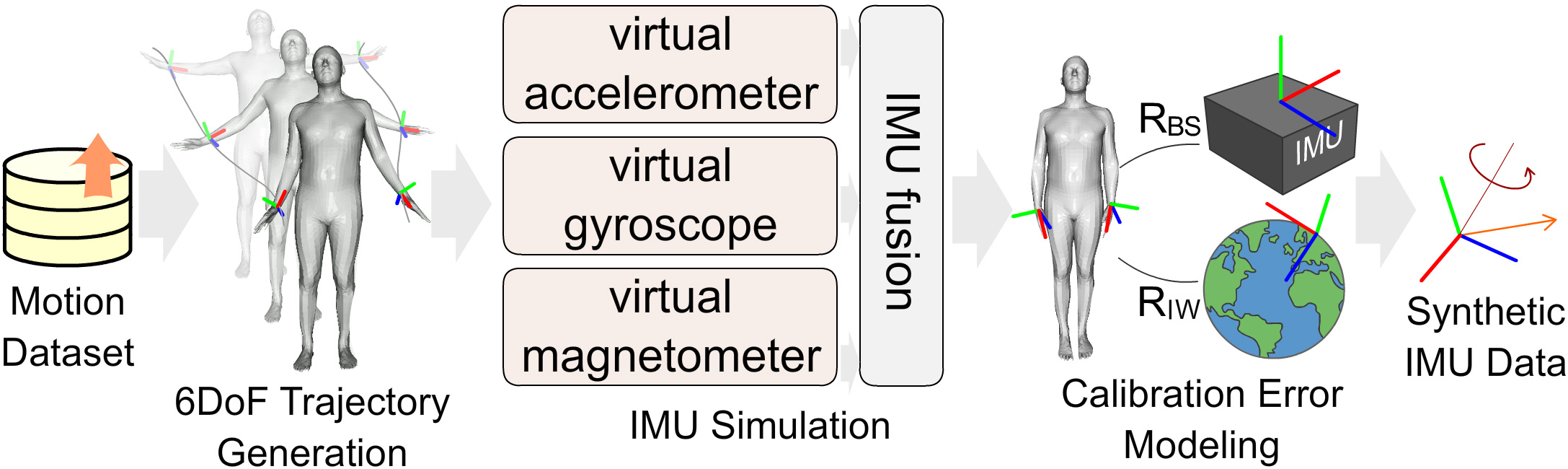}
    \caption{
        Overview of our IMU synthesis method.
        %
        %
        We first synthesize the 6DoF trajectory (position and orientation) of each IMU from the low frame-rate (60FPS) motion capture data.
        Then, we calculate the raw high frame-rate (180FPS) IMU signals including the acceleration, angular velocity, and magnetic field measurement from the trajectory.
        After adding sensor noise to the raw signals, we perform IMU fusion to get the orientation measurement.
        Finally, we simulate a T-pose calibration process, where the calibration error is added to the sensor readings.
    }
    \label{fig:methodb}
\end{figure}
\subsection{Data Synthesis}\label{sec:3.2}
Due to the lack of MoCap data containing real IMU measurements, we synthesize raw IMU signals using MoCap sequences to facilitate network training.
Previous works~\cite{DIP,PIP,TransPose} use trivial methods to generate synthetic data and the trained models work well on real data after finetuning.
However, this is not the case for this work as it relies on acceleration measurements more heavily to model fictitious forces, and the acceleration is more difficult to synthesize due to its sensitivity to the data sampling rate and the noise model.  
So in this work, we revisit the hardware and noise models to synthesize more realistic training data.
Specifically, we first synthesize IMU’s 6DoF (position and orientation) trajectories from the low frame-rate motion capture data in the world coordinate frame (Sec.~\ref{sec:3.2.1}).
Then, we solve for the low-level IMU signals in a high frame rate according to the 6DoF trajectory, add noise to the solved signals, and fuse these signals to synthesize the IMU measurements. (Sec.~\ref{sec:3.2.2}).
Finally, we model calibration errors introduced during T-pose calibration for using the synthetic measurements in the training (Sec.~\ref{sec:3.2.3}).
%
%
As examined in the experiments, our carefully generated training data leads to substantial improvements in the final results.
\par
\textit{Notations.} 
In the following, we consider each IMU individually, and all IMUs follow the same procedure. 
We start by defining several coordinate frames.
We denote the world coordinate frame of the motion capture data as $W$; in other words, all poses and translations from the datasets are expressed in the frame $W$.
We denote the IMU sensor's frame as $S$, which is a coordinate frame fixed to the IMU device.
We denote the IMU's world coordinate frame as $I$, which is a coordinate frame specified by the sensor manufacturer and is used to indicate the sensor's global orientation; in other words, the output orientation of an IMU is just its orientation with respect to the frame $I$, represented as $\boldsymbol{R}_{IS}$, following the subscript rule outlined in Sec.~\ref{sec:3.1.1}.
%
%
%
Finally, we define the human bone frame as $B$, which is a frame that is fixed to a bone segment of the human body.
This frame is useful as an IMU is always attached to a bone of the body.

%

\subsubsection{6DoF trajectory generation}\label{sec:3.2.1}
Leveraging the SMPL~\cite{SMPL} human model, we reconstruct human meshes with global translation using the MoCap sequences~\cite{AMASS}.
For each IMU, we select one vertex on the mesh of one bone as its mounting position and compute its position and orientation as:
\begin{equation}
\boldsymbol{p}_{WS}=\boldsymbol{p}_{WB}+\boldsymbol{R}_{WB}\boldsymbol{p}_{BS},
\end{equation}
\begin{equation}
\boldsymbol{R}_{WS}=\boldsymbol{R}_{WB}\boldsymbol{R}_{BS}.
\end{equation}
%
%
%
%
%
This is an ideal case, while in real capture, the sensor will slightly move around the mounting point, including translation $\delta \boldsymbol{p}_{BS}$ and rotation $\delta \boldsymbol{R}_{BS}$, which are modeled as individual random walking noises.
%
%
Considering these factors, our 6DoF trajectory of sensor position and orientation $\{\tilde{\boldsymbol{p}}_{WS}, \tilde{\boldsymbol{R}}_{WS}\}$ becomes:
\begin{equation}
    \tilde{\boldsymbol{p}}_{WS} = \boldsymbol{p}_{WB} + \boldsymbol{R}_{WB}(\boldsymbol{p}_{BS} + \delta \boldsymbol{p}_{BS}),
\end{equation}
\begin{equation}
    \tilde{\boldsymbol{R}}_{WS} = \boldsymbol{R}_{WB}\boldsymbol{R}_{BS}\delta \boldsymbol{R}_{BS}.
\end{equation}
\subsubsection{IMU Simulation}\label{sec:3.2.2}
The orientation readings from IMUs are not directly measured by the hardware, but computed from other low-level signals using the algorithm from the manufacturer.
While previous works~\cite{DIP,PIP,TransPose,TIP} directly synthesize orientation measurements, we propose to simulate the raw signals, including the accelerations, angular velocities, and magnetic field measurements, and employ the IMU fusion algorithm on our own.
Tailored for the training of neural networks, this simulation approach allows us to simulate and incorporate real-world noises with mathematical foundations.
In the following, we elaborate our method to generate the aforementioned low-level signals from the synthesized 6DoF trajectory $\{\tilde{\boldsymbol{p}}_{WS}, \tilde{\boldsymbol{R}}_{WS}\}$.
\paragraph{Virtual Accelerometer Simulation.} 
Here we synthesize acceleration $\boldsymbol{a}$.
Since the accelerometer in the IMU sensor operates at a higher frame rate than our motion capture data, we need to synthesize the accelerations at three times the frame rate (180fps) of the motion capture data (60fps) to emulate the real sensor behavior.
In this case, a finite difference method is not applicable as it does not keep the global smoothness. 
We design a novel acceleration synthesis algorithm based on energy optimization, formulated as:
%
%
\begin{equation}\label{eq:accsyn}
\begin{aligned}
    \arg\min_{\boldsymbol{a}_{ij},\boldsymbol{v}_{ij}}&\lambda_p\sum_{i=1}^{m-1}\left\|\sum_{j=1}^n\boldsymbol{v}_{ij}\Delta t+\tilde{\boldsymbol{p}}_i-\tilde{\boldsymbol{p}}_{i+1}\right\|^2\\
    +&\lambda_v\sum_{i=1}^{m-1}\left\|\sum_{j=1}^n\boldsymbol{a}_{ij}\Delta t+\boldsymbol{v}_i-\boldsymbol{v}_{i+1}\right\|^2
    +\lambda_a\sum_{i,j}\left\|\boldsymbol{a}_{ij}-\boldsymbol{a}_{ij}^{\mathrm{prev}}\right\|^2,\\
    \mathrm{where}\,\,\,\,&\boldsymbol{v}_{ij} = \boldsymbol{v}_{ij}^{\mathrm{prev}} + \boldsymbol{a}_{ij}\Delta t.
\end{aligned}
\end{equation}
Here $i\in\{1,2,\cdots,m-1\}$ is the original frame label in an $m$-frame sequence, $j \in \{1,2,3\}$ denotes the upsampled local timestamp between two original frames, $\tilde{\boldsymbol{p}}_i$, $\boldsymbol{v}_{ij}$, and $\boldsymbol{a}_{ij}$ are the global position, velocity, and acceleration of the IMU at frame $i$ or frame $ij$ (here we omit the reference frame subscripts in $(\tilde{\boldsymbol{p}}_{WS})_i$, $({\boldsymbol{v}}_{WS})_{ij}$, and $({\boldsymbol{a}}_{WS})_{ij}$ for conciseness), $\Delta t$ is the timestamp interval after upsampling, $\lambda_p$, $\lambda_v$, and $\lambda_a$ are coefficients for the energy terms, and $\cdot^{\mathrm{prev}}$ retrieves the value of the previous timestamp.
This energy optimization aims to find the high frame-rate acceleration sequence that best replicates the original low frame-rate IMU trajectory (first and second terms in Eq.~\ref{eq:accsyn}) with minimal rate of change (last term in Eq.~\ref{eq:accsyn}).
This is inspired by the research of~\cite{jerk}, which found that human motion is characterized by the minimal rate of change of acceleration.
%
%
%
\paragraph{Virtual Gyroscope Simulation.} 
Here we synthesize angular velocity $\boldsymbol{\omega}$.
Similar to the virtual accelerometer simulation, we synthesize the gyroscope's local angular velocities based on another energy optimization defined as:
\begin{equation}\label{eq:gyrsyn}
\begin{aligned}
      \arg\min_{\boldsymbol{\omega}_{ij}}&\lambda_R\sum_{i=1}^{m-1}\left\|\mathrm{Log}\left(\tilde{\boldsymbol{R}}_{i+1}^T\tilde{\boldsymbol{R}}_i\prod_{j=1}^n\mathrm{Exp}(\boldsymbol{\omega}_{ij}\Delta t)\right)\right\|^2\\
      +&\lambda_\omega\sum_{i,j}\left\|\boldsymbol{\omega}_{ij}-\boldsymbol{\omega}_{ij}^{\mathrm{prev}}\right\|^2,
\end{aligned}
\end{equation}
where $\tilde{\boldsymbol{R}}_i$ and $\boldsymbol{\omega}_{ij}$ are the orientation and the local angular velocity of the IMU at frame $i$ and frame $ij$, $\mathrm{Log}:\mathrm{SO(3)}\rightarrow\mathbb{R}^3$ maps the rotation from the Lie group to the vector space, while $\mathrm{Exp}:\mathbb{R}^3\rightarrow\mathrm{SO(3)}$ maps the rotation from the vector space to the Lie group, and $\lambda_R$ and $\lambda_\omega$ are two weight terms.
This energy optimization aims to find the high frame-rate local angular velocity sequence that best replicates the low frame-rate IMU orientations (first term in Eq.~\ref{eq:gyrsyn}) with minimal rate of change (last term in Eq.~\ref{eq:gyrsyn}).
%
%
\paragraph{Virtual Magnetometer Simulation.} 
The real magnetometer typically works at the same frame rate as our motion capture data, so upsampling is not necessary.
We assume the x-axis is the north direction and transform it into the sensor frame to obtain the magnetic field measurement $\boldsymbol{m}_i$ for each frame.
%
\paragraph{IMU fusion.} 
In this stage, we fuse the synthesized accelerometer, gyroscope, and magnetometer measurements to estimate the sensor's orientations.
Similar to~\cite{vinsmono}, we consider the measurement noise as sensor bias and additive noise, modeled by random walks and Gaussian white noise, respectively.
These noises are sampled and added to the raw sensor signals to get the noisy measurements denoted as $\tilde{\boldsymbol{a}}_{ij}, \tilde{\boldsymbol{\omega}}_{ij}, \tilde{\boldsymbol{m}}_{i}$.
Then, we implement an advanced sensor fusion algorithm utilizing the error-state Kalman Filter~\cite{ESKF}, bolstered by the inclusion of Zero Velocity Update (ZUPT)~\cite{zupt} detection for state updates.
%
%
Readers can refer to~\cite{ESKF} for the algorithm.
After the sensor fusion, we obtain IMU orientation $\boldsymbol{R}_{IS}$ in the IMU world frame defined by our algorithm, formulated as:
\begin{equation}
    (\boldsymbol{R}_{IS})_{ij} = \mathrm{ESKF}(\tilde{\boldsymbol{a}}_{ij}, \tilde{\boldsymbol{\omega}}_{ij}, \tilde{\boldsymbol{m}}_{i}).
\end{equation}
\subsubsection{Calibration Error Modeling}\label{sec:3.2.3}
To train the mocap network, we need to transform the simulated IMU orientation $\boldsymbol{R}_{IS}$ (sensor orientation in the IMU world frame) to the actual bone orientation $\boldsymbol{R}_{WB}$ (bone orientation in the world frame), which is formulated as:
\begin{equation}\label{eq:calib}
    \boldsymbol{R}_{WB} = \boldsymbol{R}_{IW}^T\boldsymbol{R}_{IS}\boldsymbol{R}_{BS}^T.
\end{equation}
For synthetic data, the two matrices are unknown: $\boldsymbol{R}_{BS}$ indicating how each sensor is placed relative to the body, and $\boldsymbol{R}_{IW}$ indicating how the IMU global frame is rotated with respect to the motion capture world frame (see Fig.~\ref{fig:methodb}).
For real data, these two matrices are typically determined through a T-pose calibration in practice, where the subject stands in a known T pose and the two matrices are estimated~\cite{TransPose,DIP}.
However, as the subject cannot perform a perfect T-pose, the calibration step contains errors.
So in this work, during data synthesis, we model such calibration errors by perturbing the two matrices $\boldsymbol{R}_{IW}$ and $\boldsymbol{R}_{BS}$ with random noise before applying Eq.~\ref{eq:calib}.
%
%

%
\begin{figure}[t]
    \centering
    \includegraphics[width=\linewidth]{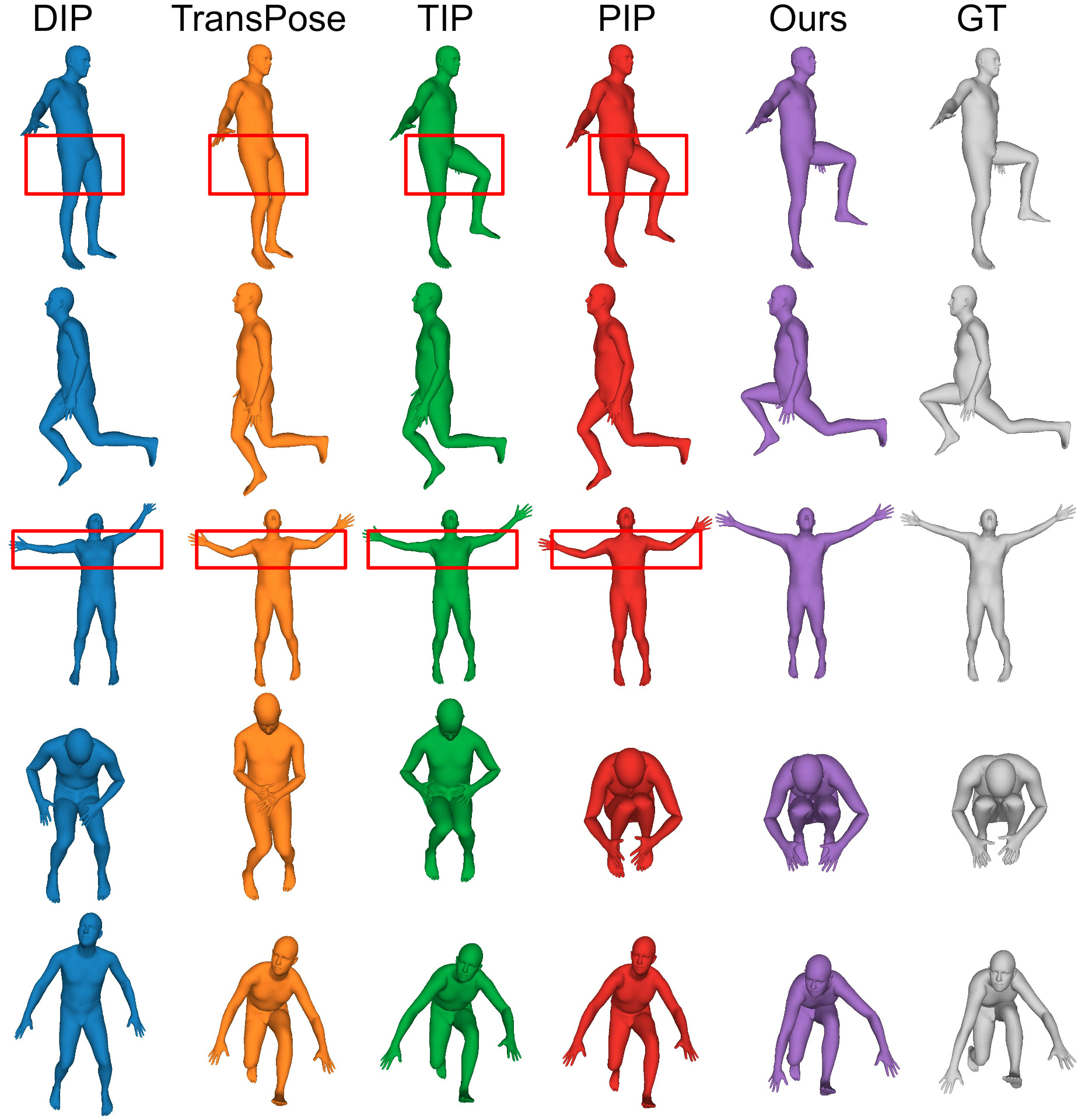}
    \caption{Qualitative comparisons with prior works. The examples are picked from the TotalCapture~\cite{TotalCapture} dataset. }
    \label{fig:posecmp}
\end{figure}

\begin{figure*}[h]
    \centering
    \subfigure[Qualitative evaluation on fictitious acceleration.]{\includegraphics[width=0.47\textwidth]{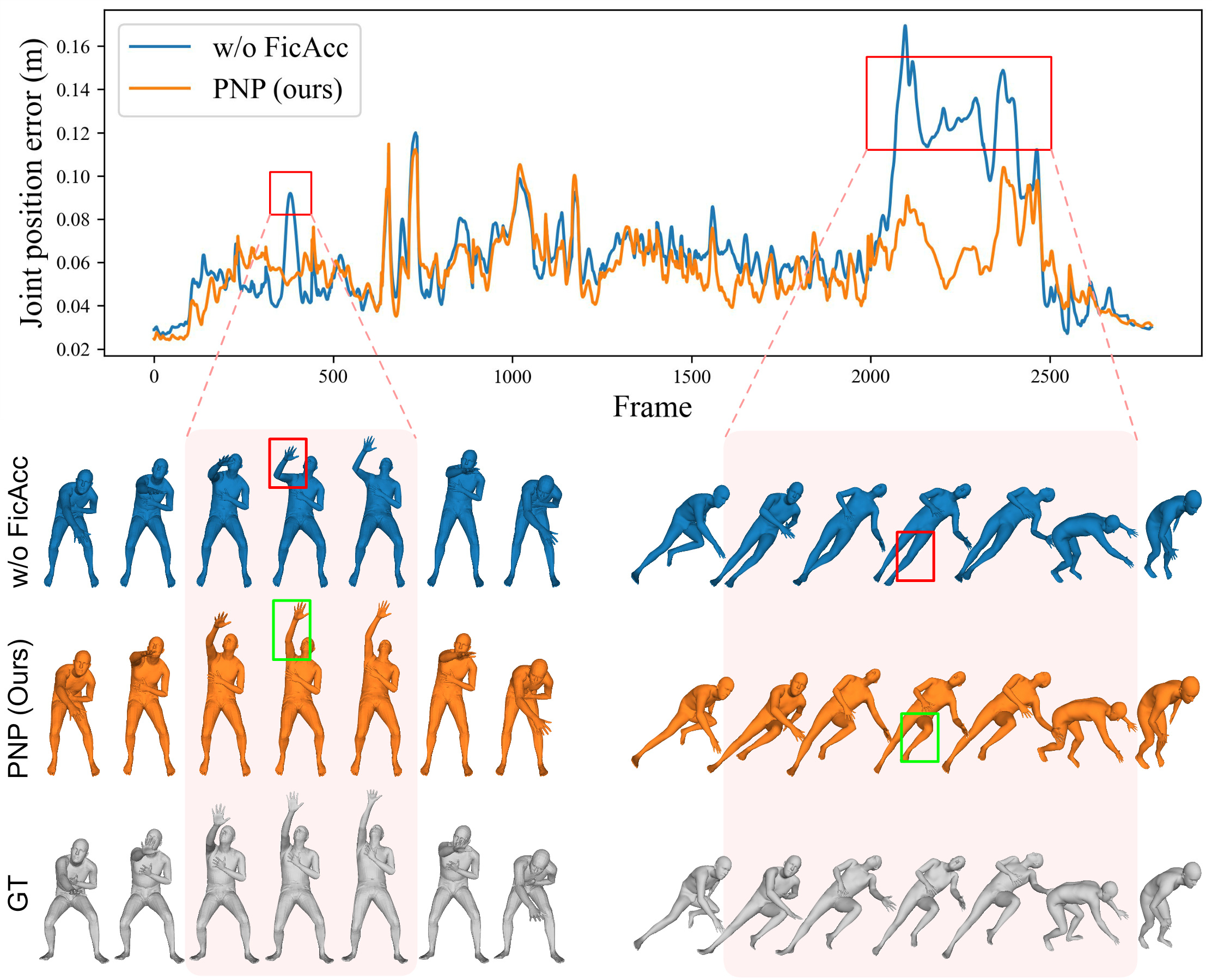}}\hfill
    \subfigure[Qualitative evaluation on the proposed IMU synthesizing method.]{\includegraphics[width=0.47\textwidth]{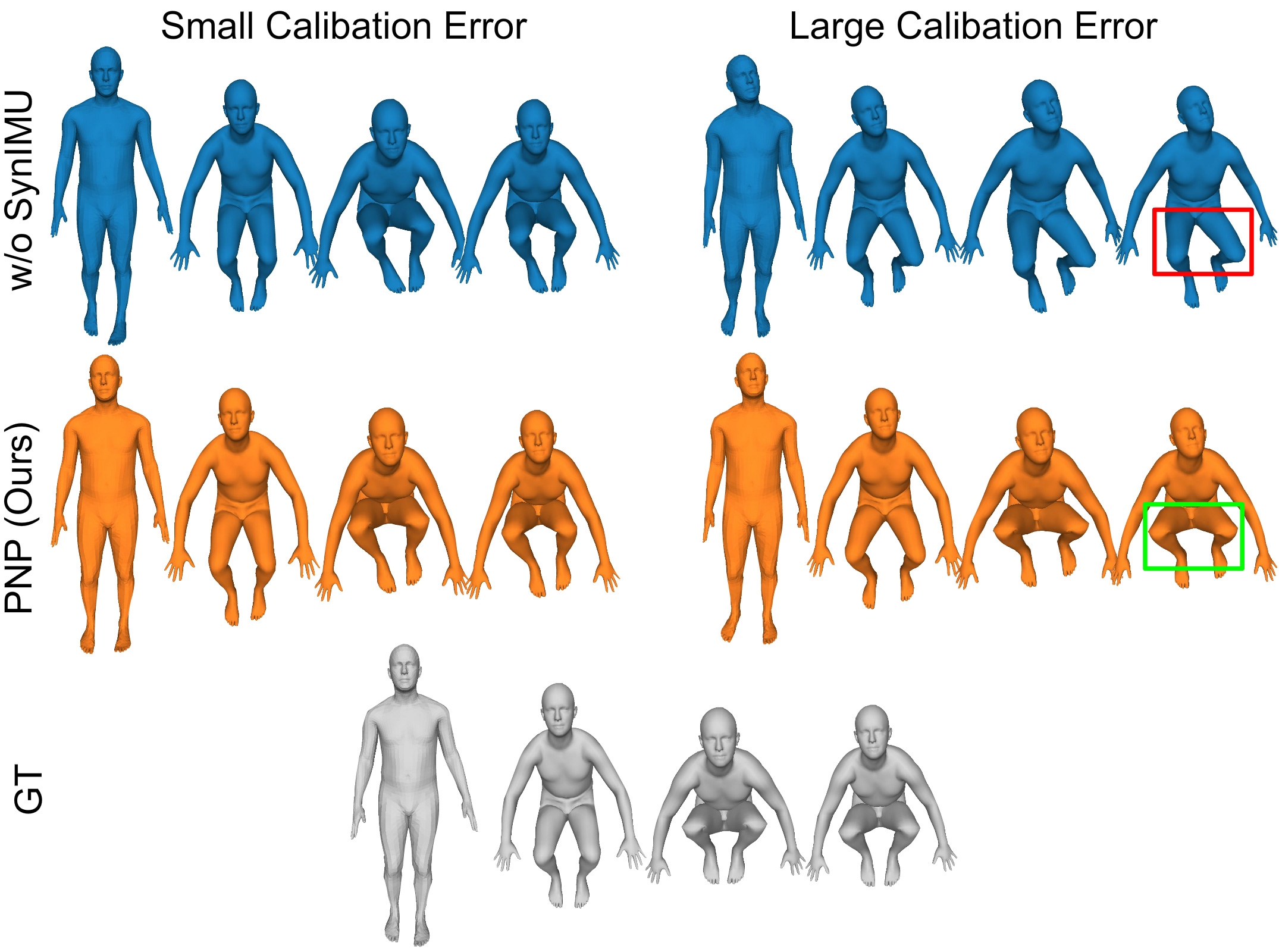}}
    \caption{Qualitative evaluation on the proposed (a) fictitious acceleration and (b) the IMU synthesizing method. (a) We plot the joint position error in a sequence from the TotalCapture dataset and compare the reconstructed motions at two selected time intervals. (b) We visualize the predicted motion for the same sequence under both small and large calibration errors (DIP-calibration vs. official calibration).}
    \label{fig:abl}
\end{figure*}

\section{Experiments}
In this section, we present the implementation details (Sec.~\ref{sec:4.1}).
We then compare our method with state-of-the-art works in motion capture from sparse IMUs (Sec.~\ref{sec:4.2}) and evaluate the key contributions of our method (Sec.~\ref{sec:4.3}).
Finally, we discuss our limitations (Sec.~\ref{sec:4.4}). 
Please also find more results in the supplemental video.
\subsection{Implementation Details}\label{sec:4.1}
\paragraph{Networks}
Our method incorporates 6 neural networks, including 1 fully connected network for the fictitious force estimator, and 5 recurrent networks for local pose and global motion estimation following~\cite{PIP}.
The fictitious force estimator contains 4 layers with a hidden width of 512, activated by ReLU, and optimized by the Adam optimizer~\cite{Adam}. 
%
%
%
%
%
%
%
\paragraph{Data Synthesis}
In 6DoF trajectory generation, we model the sensor sliding error $\delta \boldsymbol{p}_{BS}$ as a random walk beginning with an error expectation of $10^{-2}\mathrm{m}$ and progressing at a speed of $10^{-3}\mathrm{m/s}$.
The rotation error $\delta \boldsymbol{R}_{BS}$ is modeled as a random walk beginning with zero error and progressing at a speed of $10^{-2}\mathrm{rad/s}$.
These values are set based on experience. 
For the energy optimization in the IMU simulation, we set $\lambda_p=1$, $\lambda_v=0.5$, $\lambda_a=1.3$, $\lambda_R=1$, and $\lambda_\omega=1$.
In the calibration error modeling, we set the perturbation to $\boldsymbol{R}_{IM}$ and $\boldsymbol{R}_{BS}$ with a mean of 0.01 and 0.1 radians, respectively.
The IMU noise parameters are set following~\cite {ORBSLAM3}\footnote{\hl{The used parameters are compatible with most mocap IMUs. For specific requirements, one can determine the noise parameter of their IMUs using tools like Kalibr (\url{https://github.com/ethz-asl/kalibr}) or refer to the manufacturer's hardware specifications.}}.
\paragraph{Hardware and Performance}
Our method runs in real time at 60fps on a laptop with Intel(R) Core(TM) i7-12700H CPU without GPU.
For live demo, we use PN Lab sensors from Noitom~\cite{Noitom}.
Our framework is implemented in Pytorch~\cite{Pytorch} and the physics-based optimization is implemented using Rigid Body Dynamic Library (RBDL)~\cite{RBDL}.  
The IMU fusion algorithm is implemented in C++, and the optimization is solved with sparse least square solvers~\cite{lsqr}.
%
%
\paragraph{Datasets}
The datasets comprise DIP-IMU~\cite{DIP}, TotalCapture~\cite{TotalCapture}, and AMASS~\cite{AMASS}.
We follow previous works~\cite{PIP,TIP} to split the train and test sets. 
%
%
Notably, there are two different calibrations for TotalCapture.
The official calibration (denoted as \textit{Official Calibration}) provided in~\cite{TotalCapture} has a larger calibration error of 12.1 degrees, while the calibration from \cite{DIP} (denoted as \textit{DIP Calibration}) on the same data has a lower calibration error of 8.6 degrees.
As the two versions differ only in calibration, we use them to examine the methods' robustness against calibration.
\begin{table}[t]
\centering
\caption{%
    Quantitative comparison results with prior works. We adopt two versions of TotalCapture: the Official Calibration has a larger calibration error in the IMU input than the DIP Calibration version.
}
\resizebox{\linewidth}{!}{
\begin{tabular}{cccccc}
\toprule
Method     & SIP Error      & Ang Error      & Pos Error     & Mesh Error     & Jitter        \\ 
\hline \multicolumn{6}{c}{TotalCapture (Official Calibration)} \\ \hline
DIP        & 18.73          & 17.57          & 9.47          & 11.33          & -             \\
TransPose  & 18.12          & 14.91          & 7.10          & 8.09           & 1.95          \\
TIP        & 15.62          & 14.45          & 6.76          & 7.79           & 1.74          \\
PIP        & 14.52          & 13.85          & 6.22          & 7.21           & \textbf{0.21} \\ 
PNP (Ours) & \textbf{11.35} & \textbf{11.10} & \textbf{4.89} & \textbf{5.60}  & 0.27          \\ 
\hline \multicolumn{6}{c}{TotalCapture (DIP Calibration)} \\ \hline
DIP        & 18.62          & 17.22          & 9.42          & 11.22          & -             \\
TransPose  & 16.58          & 12.89          & 6.55          & 7.42           & 1.87          \\
TIP        & 13.22          & 12.30          & 5.81          & 6.80           & 1.69          \\
PIP        & 12.93          & 12.04          & 5.61          & 6.51           & \textbf{0.20} \\ 
PNP (Ours) & \textbf{10.89} & \textbf{10.45} & \textbf{4.74} & \textbf{5.45}  & 0.26          \\ 
\hline \multicolumn{6}{c}{DIP-IMU} \\ \hline
DIP        & 17.35          & 15.36          & 7.59          & 9.05           & -             \\
TransPose  & 17.06          & 8.86           & 6.03          & 7.17           & 1.11          \\
TIP        & 16.90          & 9.07           & 5.63          & 6.62           & 1.56          \\
PIP        & 15.33          & 8.78           & 5.12          & 6.02           & \textbf{0.17} \\ 
PNP (Ours) & \textbf{13.71} & \textbf{8.75} & \textbf{4.97} & \textbf{5.77}   & \textbf{0.17} \\ 
\bottomrule
\end{tabular}\label{tab:posecmp}
}
\end{table}
\begin{figure}
    \centering
    \includegraphics[width=\linewidth]{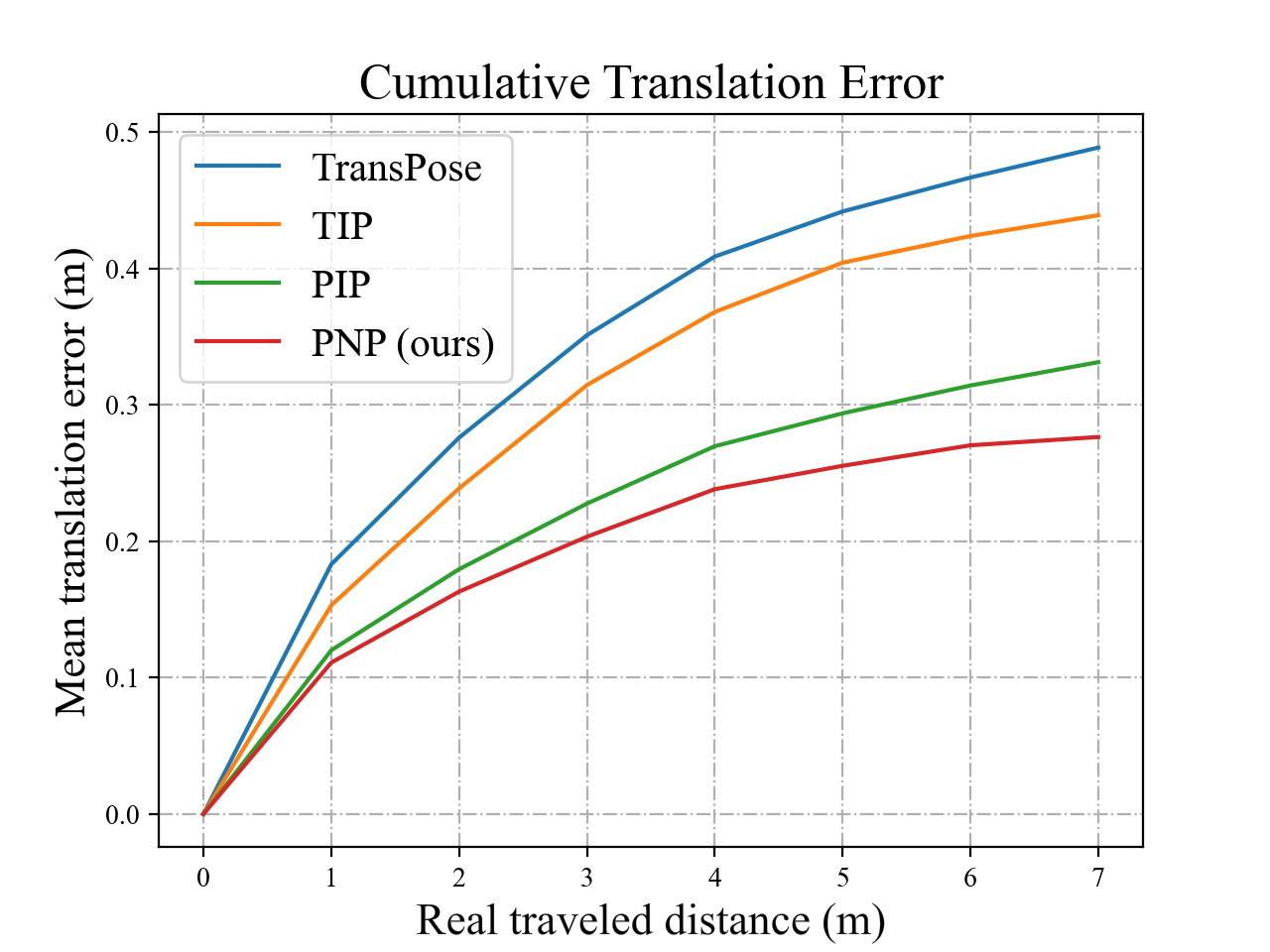}
    \caption{Translation comparisons with prior works. We plot the global position error accumulation curve with respect to the real traveled distance. A lower curve indicates smaller drift.}
    \label{fig:trancmp}
\end{figure}
\subsection{Comparisons}\label{sec:4.2}
We compare our method with the state-of-the-art works in motion capture from sparse IMUs, including DIP~\cite{DIP}, TransPose~\cite{TransPose}, TIP~\cite{TIP}, and PIP~\cite{PIP}.
When evaluating local pose accuracy, we align the root joint position and orientation with the ground truth, and use the same metrics as in~\cite{TransPose,PIP}, including:
\begin{itemize}
    \item SIP Error (${}^{\circ}$): the global rotation error of hips and shoulders.
    \item Angular Error (${}^{\circ}$): the global rotation error of all joints.
    \item Positional Error ($\mathrm{cm}$): the position error of all joints.
    \item Mesh Error ($\mathrm{cm}$): the vertex error of the posed SMPL meshes.
    \item Jitter ($\mathrm{10^3m/s^3}$): the average jerk of all joints \textit{w.r.t} the world.
\end{itemize}
The pose comparison results are presented in Tab.~\ref{tab:posecmp}.
Our method consistently outperforms previous works in pose accuracy, achieving a reduction of 19\% in mesh error in the TotalCapture dataset and 4\% in the DIP-IMU dataset. 
As all these methods are trained on the DIP-IMU dataset, our significant improvement on the TotalCapture dataset demonstrates that our method generalizes better, thanks to our carefully synthesized training data.
While our method has a slightly larger jitter than PIP, this does not mean our system is shaky, but rather more sensitive to accelerations and captures the movements more accurately.
As shown in Fig.~\ref{fig:posecmp} and the supplementary video, our method faithfully reconstructs the large movements, while other methods are over-smooth.
This improvement comes from our utilization of \textit{corrected} acceleration in a non-inertial frame.
\par
We compare the estimated translation with prior works in Fig.~\ref{fig:trancmp}.
Our method shows the lowest global position drift compared to prior works, as the algorithm for translation estimation heavily depends on the estimated pose. 
%
%
%
%
%
%
%
%
%

\begin{table}[]
\caption{Evaluation on fictitious acceleration (FicAcc) and IMU synthesis (SynIMU) with 2 different calibrations: the official calibration (higher calibration error) and the DIP calibration (lower calibration error).}
\label{tab:abl2}
\resizebox{\linewidth}{!}{

\begin{tabular}{cccccc}
\toprule
Method & SIP Error      & Ang Error      & Pos Error     & Mesh Error     & Jitter        \\
\hline \multicolumn{6}{c}{TotalCapture (official calibration)} \\ \hline
w/o SynIMU      & 13.12          & 13.70          & 5.93          & 6.82           & 0.28          \\
w/o FicAcc      & 11.43          & 11.16          & 4.93          & 5.68           & \textbf{0.27} \\
\hl{w/o normalization}	& \hl{11.71}          & \hl{11.31}          & \hl{5.05}          & \hl{5.81}           & \hl{\textbf{0.27}} \\
\hl{FicAcc by Eq. 1}	& \hl{11.67}          & \hl{11.26}          & \hl{5.11}          & \hl{5.81}           & \hl{\textbf{0.27}} \\
PNP (Ours)      & \textbf{11.35} & \textbf{11.10} & \textbf{4.89} & \textbf{5.60}  & \textbf{0.27} \\
\hline \multicolumn{6}{c}{TotalCapture (DIP calibration)} \\ \hline
w/o SynIMU      & 11.28          & 11.38          & 4.89          & 5.64           & 0.27          \\
w/o FicAcc      & \textbf{10.82} & 10.58          & 4.78          & 5.53           & \textbf{0.26} \\
\hl{w/o normalization} & \hl{11.31}          & \hl{10.70}          & \hl{4.95}          & \hl{5.72}           & \hl{\textbf{0.26}} \\
\hl{FicAcc by Eq. 1} & \hl{10.94}          & \hl{10.51}          & \hl{4.86}          & \hl{5.56}           & \hl{\textbf{0.26}} \\
PNP (Ours)      & 10.89          & \textbf{10.45} & \textbf{4.74} & \textbf{5.45}  & \textbf{0.26} \\
\bottomrule
\end{tabular}}
\end{table}



%

%
%
%
%
\subsection{Evaluations}\label{sec:4.3}
\begin{figure}[t]
    \centering
    \includegraphics[width=\linewidth]{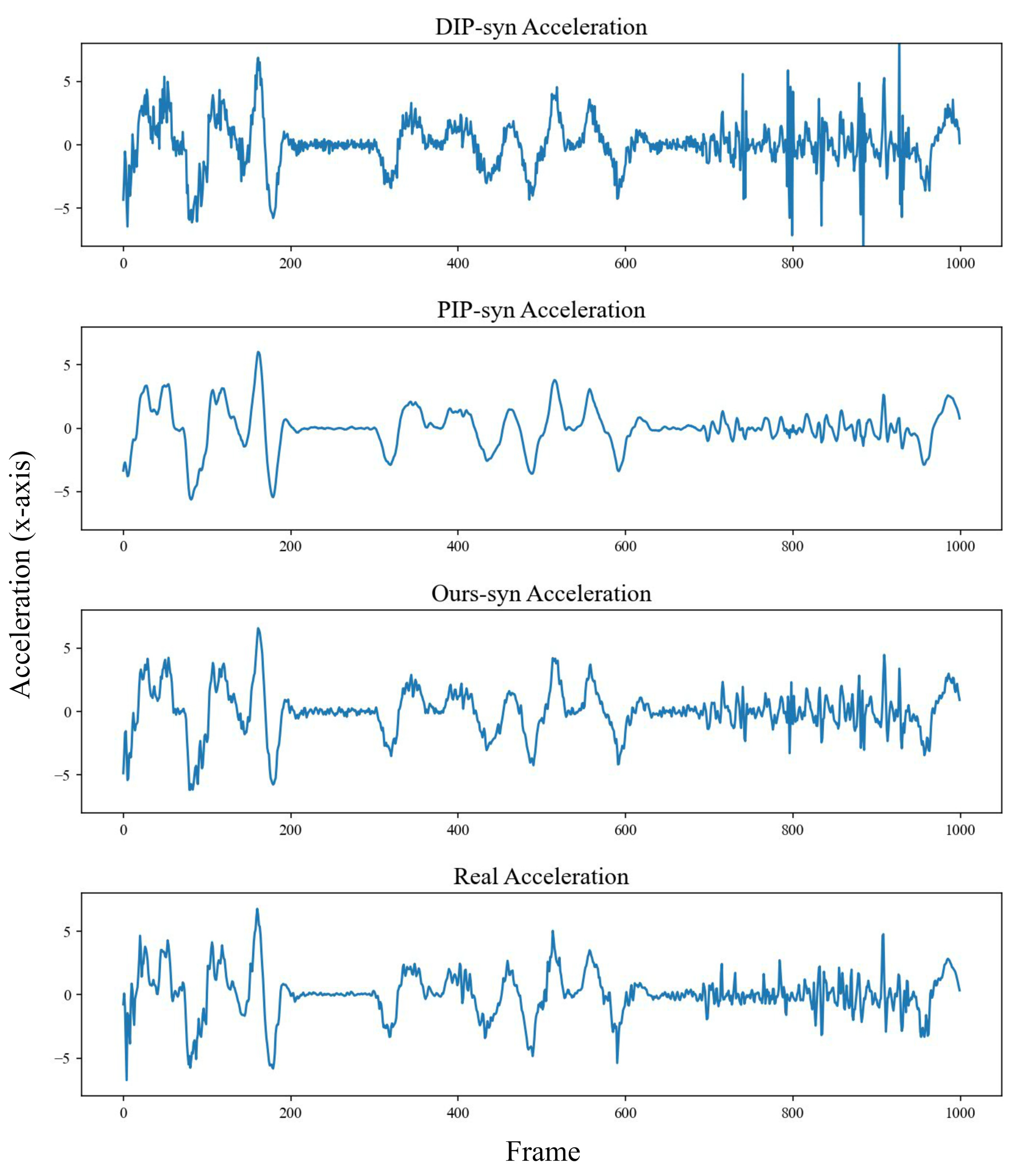}
    \caption{Comparisons of synthetic acceleration measurements using different methods, including the method in DIP~\cite{DIP}, PIP~\cite{PIP}, and the proposed method. The bottom figure shows the IMU accelerations of real sensors obtained from the dataset. The acceleration synthesis method in DIP results in excessive jitter, while PIP's method tends to oversmooth the acceleration. In contrast, our synthetic acceleration most accurately reflects the behavior of the real sensor.}
    \label{fig:acccmp}
\end{figure}

In this section, we evaluate our key contributions: the fictitious force for acceleration correction and the novel IMU synthesis method.
In the first ablation study (w/o SynIMU), we train our model using the synthetic data from \cite{PIP}, which does not consider the low-level IMU signals nor the calibration error.
In the second evaluation (w/o FicAcc), we exclude the fictitious force estimation, meaning the leaf accelerations are naively projected to the root frame and are subtracted by the root acceleration as similar to~\cite{PIP}.
\hl{In the third evaluation (w/o normalization), the accelerations are transformed to the root frame without subtracting the root acceleration.
In the fourth evaluation (FicAcc by Eq. 1), we replace the neural fictitious acceleration estimator with the analytical calculation using Eq.~\ref{eq:fictitious-force}.}
These variants are tested on the TotalCapture dataset using both the official calibration and DIP-calibration.
As reported in Tab.~\ref{tab:abl2}, our full method is generally better.
The numerical gap between ours and \textit{w/o FicAcc} is less obvious.
%
%
\hl{This is because modeling fictitious force is particularly beneficial for handling ambiguous movements with large accelerations and minimal orientation changes, which only take a small portion of the dataset, diluting the quantitative numbers.}
We demonstrate it through Fig.~\ref{fig:abl} (a), where we select a sequence from the test data.
While most of the time the two methods have identical errors, the difference becomes substantial for a few ambiguous motions where acceleration is necessary for pose estimation, \textit{e.g.}, raising arms and legs as highlighted in the figure.
We further note that the variant using the vanilla IMU synthesis method (w/o SynIMU) demonstrates a huge gap under different calibrations in Tab.~\ref{tab:abl2}, while our method performs consistently well.
The qualitative comparison is demonstrated in Fig.~\ref{fig:abl} (b).
Our full method is robust to calibration errors as we explicitly model them when producing training data.
Additionally, we directly compare the generated acceleration by different IMU synthesis methods in Fig.~\ref{fig:acccmp}. 
Our synthesized acceleration closely resembles the real sensor measurement compared with previous methods.
\subsection{Limitations}\label{sec:4.4}
%
%
Our IMU synthesis technique assumes a uniform and constant magnetic field.
However, magnetic disturbances are common in real world, and our method is less robust to magnetic changes.
This work only focuses on pose estimation and assumes the captured subject is in an average shape, while in real world the pose and shape are coupled in the IMU measurements.
%
%
Finally, we only consider a flat ground and do not support walking up- or downstairs.
%
%
%
%
%

%
\section{Conclusion}
This work proposes a novel method for real-time motion capture from sparse IMUs.
To the best of our knowledge, it is the first work to model the non-inertial effects, whose impact on the acceleration measurements is non-negligible.
By training an auto-regressive network to estimate the fictitious forces and modulate the IMU accelerations, our method better exploits the acceleration information to capture challenging poses.
We further propose an IMU synthesis method that fully considers the real IMU hardware noises and calibration errors to generate more realistic data.
By narrowing the gap between the synthetic and real, our model generalizes better.

\begin{acks}
The authors would like to thank Notiom for the extensive support on inertial sensors, and Liangdi Ma, Yunzhe Shao, Shaohua Pan for the help on live demos. This work was supported by the National Key R\&D Program of China (2023YFC3305600, 2018YFA0704000), the NSFC (No.62021002), and the Key Research and Development Project of Tibet Autonomous Region (XZ202101ZY0019G). This work was also supported by THUIBCS, Tsinghua University, and BLBCI, Beijing Municipal Education Commission. Feng Xu is the corresponding author.
\end{acks}

\bibliographystyle{ACM-Reference-Format}
\bibliography{cv}


\begin{thebibliography}{50}


\ifx \showCODEN    \undefined \def \showCODEN     #1{\unskip}     \fi
\ifx \showDOI      \undefined \def \showDOI       #1{#1}\fi
\ifx \showISBNx    \undefined \def \showISBNx     #1{\unskip}     \fi
\ifx \showISBNxiii \undefined \def \showISBNxiii  #1{\unskip}     \fi
\ifx \showISSN     \undefined \def \showISSN      #1{\unskip}     \fi
\ifx \showLCCN     \undefined \def \showLCCN      #1{\unskip}     \fi
\ifx \shownote     \undefined \def \shownote      #1{#1}          \fi
\ifx \showarticletitle \undefined \def \showarticletitle #1{#1}   \fi
\ifx \showURL      \undefined \def \showURL       {\relax}        \fi
\providecommand\bibfield[2]{#2}
\providecommand\bibinfo[2]{#2}
\providecommand\natexlab[1]{#1}
\providecommand\showeprint[2][]{arXiv:#2}

\bibitem[Ahuja et~al\mbox{.}(2021)]%
        {ahuja2021coolmoves}
\bibfield{author}{\bibinfo{person}{Karan Ahuja}, \bibinfo{person}{Eyal Ofek}, \bibinfo{person}{Mar Gonzalez-Franco}, \bibinfo{person}{Christian Holz}, {and} \bibinfo{person}{Andrew~D Wilson}.} \bibinfo{year}{2021}\natexlab{}.
\newblock \showarticletitle{Coolmoves: User motion accentuation in virtual reality}.
\newblock \bibinfo{journal}{\emph{Proceedings of the ACM on Interactive, Mobile, Wearable and Ubiquitous Technologies}} \bibinfo{volume}{5}, \bibinfo{number}{2} (\bibinfo{year}{2021}), \bibinfo{pages}{1--23}.
\newblock


\bibitem[Aliakbarian et~al\mbox{.}(2022)]%
        {aliakbarian2022flag}
\bibfield{author}{\bibinfo{person}{Sadegh Aliakbarian}, \bibinfo{person}{Pashmina Cameron}, \bibinfo{person}{Federica Bogo}, \bibinfo{person}{Andrew Fitzgibbon}, {and} \bibinfo{person}{Thomas~J Cashman}.} \bibinfo{year}{2022}\natexlab{}.
\newblock \showarticletitle{Flag: Flow-based 3d avatar generation from sparse observations}. In \bibinfo{booktitle}{\emph{Proceedings of the IEEE/CVF Conference on Computer Vision and Pattern Recognition}}. \bibinfo{pages}{13253--13262}.
\newblock


\bibitem[Aliakbarian et~al\mbox{.}(2023)]%
        {Aliakbarian_2023_ICCV}
\bibfield{author}{\bibinfo{person}{Sadegh Aliakbarian}, \bibinfo{person}{Fatemeh Saleh}, \bibinfo{person}{David Collier}, \bibinfo{person}{Pashmina Cameron}, {and} \bibinfo{person}{Darren Cosker}.} \bibinfo{year}{2023}\natexlab{}.
\newblock \showarticletitle{HMD-NeMo: Online 3D Avatar Motion Generation From Sparse Observations}. In \bibinfo{booktitle}{\emph{Proceedings of the IEEE/CVF International Conference on Computer Vision (ICCV)}}. \bibinfo{pages}{9622--9631}.
\newblock


\bibitem[Campos et~al\mbox{.}(2021)]%
        {ORBSLAM3}
\bibfield{author}{\bibinfo{person}{Carlos Campos}, \bibinfo{person}{Richard Elvira}, \bibinfo{person}{Juan~J. G\'omez}, \bibinfo{person}{Jos\'e M.~M. Montiel}, {and} \bibinfo{person}{Juan~D. Tard\'os}.} \bibinfo{year}{2021}\natexlab{}.
\newblock \showarticletitle{{ORB-SLAM3}: An Accurate Open-Source Library for Visual, Visual-Inertial and Multi-Map {SLAM}}.
\newblock \bibinfo{journal}{\emph{IEEE Transactions on Robotics}} \bibinfo{volume}{37}, \bibinfo{number}{6} (\bibinfo{year}{2021}), \bibinfo{pages}{1874--1890}.
\newblock


\bibitem[Castillo et~al\mbox{.}(2023)]%
        {castillo2023bodiffusion}
\bibfield{author}{\bibinfo{person}{Angela Castillo}, \bibinfo{person}{Maria Escobar}, \bibinfo{person}{Guillaume Jeanneret}, \bibinfo{person}{Albert Pumarola}, \bibinfo{person}{Pablo Arbel{\'a}ez}, \bibinfo{person}{Ali Thabet}, {and} \bibinfo{person}{Artsiom Sanakoyeu}.} \bibinfo{year}{2023}\natexlab{}.
\newblock \showarticletitle{BoDiffusion: Diffusing Sparse Observations for Full-Body Human Motion Synthesis}.
\newblock \bibinfo{journal}{\emph{arXiv preprint arXiv:2304.11118}} (\bibinfo{year}{2023}).
\newblock


\bibitem[Dittadi et~al\mbox{.}(2021)]%
        {dittadi2021full}
\bibfield{author}{\bibinfo{person}{Andrea Dittadi}, \bibinfo{person}{Sebastian Dziadzio}, \bibinfo{person}{Darren Cosker}, \bibinfo{person}{Ben Lundell}, \bibinfo{person}{Thomas~J Cashman}, {and} \bibinfo{person}{Jamie Shotton}.} \bibinfo{year}{2021}\natexlab{}.
\newblock \showarticletitle{Full-body motion from a single head-mounted device: Generating smpl poses from partial observations}. In \bibinfo{booktitle}{\emph{Proceedings of the IEEE/CVF International Conference on Computer Vision}}. \bibinfo{pages}{11687--11697}.
\newblock


\bibitem[Du et~al\mbox{.}(2023)]%
        {agrol}
\bibfield{author}{\bibinfo{person}{Yuming Du}, \bibinfo{person}{Robin Kips}, \bibinfo{person}{Albert Pumarola}, \bibinfo{person}{Sebastian Starke}, \bibinfo{person}{Ali Thabet}, {and} \bibinfo{person}{Artsiom Sanakoyeu}.} \bibinfo{year}{2023}\natexlab{}.
\newblock \showarticletitle{Avatars grow legs: Generating smooth human motion from sparse tracking inputs with diffusion model}. In \bibinfo{booktitle}{\emph{Proceedings of the IEEE/CVF Conference on Computer Vision and Pattern Recognition}}. \bibinfo{pages}{481--490}.
\newblock


\bibitem[Felis(2017)]%
        {RBDL}
\bibfield{author}{\bibinfo{person}{Martin Felis}.} \bibinfo{year}{2017}\natexlab{}.
\newblock \showarticletitle{RBDL: an efficient rigid-body dynamics library using recursive algorithms}.
\newblock \bibinfo{journal}{\emph{Autonomous Robots}}  \bibinfo{volume}{41} (\bibinfo{date}{02} \bibinfo{year}{2017}).
\newblock


\bibitem[Flash and Hogan(1985)]%
        {jerk}
\bibfield{author}{\bibinfo{person}{Tamar Flash} {and} \bibinfo{person}{Neville Hogan}.} \bibinfo{year}{1985}\natexlab{}.
\newblock \showarticletitle{The Coordination of Arm Movements: An Experimentally Confirmed Mathematical Model}.
\newblock \bibinfo{journal}{\emph{The Journal of neuroscience : the official journal of the Society for Neuroscience}}  \bibinfo{volume}{5} (\bibinfo{date}{08} \bibinfo{year}{1985}).
\newblock


\bibitem[Hochreiter and Schmidhuber(1997)]%
        {LSTM}
\bibfield{author}{\bibinfo{person}{Sepp Hochreiter} {and} \bibinfo{person}{Jürgen Schmidhuber}.} \bibinfo{year}{1997}\natexlab{}.
\newblock \showarticletitle{Long Short-term Memory}.
\newblock \bibinfo{journal}{\emph{Neural computation}}  \bibinfo{volume}{9} (\bibinfo{date}{12} \bibinfo{year}{1997}).
\newblock


\bibitem[Huang et~al\mbox{.}(2018)]%
        {DIP}
\bibfield{author}{\bibinfo{person}{Yinghao Huang}, \bibinfo{person}{Manuel Kaufmann}, \bibinfo{person}{Emre Aksan}, \bibinfo{person}{Michael~J. Black}, \bibinfo{person}{Otmar Hilliges}, {and} \bibinfo{person}{Gerard Pons-Moll}.} \bibinfo{year}{2018}\natexlab{}.
\newblock \showarticletitle{Deep Inertial Poser Learning to Reconstruct Human Pose from SparseInertial Measurements in Real Time}.
\newblock \bibinfo{journal}{\emph{ACM Transactions on Graphics, (Proc. SIGGRAPH Asia)}}  \bibinfo{volume}{37} (\bibinfo{date}{nov} \bibinfo{year}{2018}).
\newblock


\bibitem[Jiang et~al\mbox{.}(2023)]%
        {jiang2023egoposer}
\bibfield{author}{\bibinfo{person}{Jiaxi Jiang}, \bibinfo{person}{Paul Streli}, \bibinfo{person}{Manuel Meier}, \bibinfo{person}{Andreas Fender}, {and} \bibinfo{person}{Christian Holz}.} \bibinfo{year}{2023}\natexlab{}.
\newblock \showarticletitle{EgoPoser: Robust Real-Time Ego-Body Pose Estimation in Large Scenes}.
\newblock \bibinfo{journal}{\emph{arXiv preprint arXiv:2308.06493}} (\bibinfo{year}{2023}).
\newblock


\bibitem[Jiang et~al\mbox{.}(2022a)]%
        {avatarposer}
\bibfield{author}{\bibinfo{person}{Jiaxi Jiang}, \bibinfo{person}{Paul Streli}, \bibinfo{person}{Huajian Qiu}, \bibinfo{person}{Andreas Fender}, \bibinfo{person}{Larissa Laich}, \bibinfo{person}{Patrick Snape}, {and} \bibinfo{person}{Christian Holz}.} \bibinfo{year}{2022}\natexlab{a}.
\newblock \showarticletitle{Avatarposer: Articulated full-body pose tracking from sparse motion sensing}. In \bibinfo{booktitle}{\emph{Computer Vision--ECCV 2022: 17th European Conference, Tel Aviv, Israel, October 23--27, 2022, Proceedings, Part V}}. Springer, \bibinfo{pages}{443--460}.
\newblock


\bibitem[Jiang et~al\mbox{.}(2022b)]%
        {TIP}
\bibfield{author}{\bibinfo{person}{Yifeng Jiang}, \bibinfo{person}{Yuting Ye}, \bibinfo{person}{Deepak Gopinath}, \bibinfo{person}{Jungdam Won}, \bibinfo{person}{Alexander~W. Winkler}, {and} \bibinfo{person}{C.~Karen Liu}.} \bibinfo{year}{2022}\natexlab{b}.
\newblock \showarticletitle{Transformer Inertial Poser: Real-Time Human Motion Reconstruction from Sparse IMUs with Simultaneous Terrain Generation}. In \bibinfo{booktitle}{\emph{SIGGRAPH Asia 2022 Conference Papers}}.
\newblock


\bibitem[Kingma and Ba(2014)]%
        {Adam}
\bibfield{author}{\bibinfo{person}{Diederik Kingma} {and} \bibinfo{person}{Jimmy Ba}.} \bibinfo{year}{2014}\natexlab{}.
\newblock \showarticletitle{Adam: A Method for Stochastic Optimization}.
\newblock \bibinfo{journal}{\emph{International Conference on Learning Representations}} (\bibinfo{date}{12} \bibinfo{year}{2014}).
\newblock


\bibitem[Kwon et~al\mbox{.}(2020)]%
        {imutube}
\bibfield{author}{\bibinfo{person}{Hyeokhyen Kwon}, \bibinfo{person}{Catherine Tong}, \bibinfo{person}{Harish Haresamudram}, \bibinfo{person}{Yan Gao}, \bibinfo{person}{Gregory~D Abowd}, \bibinfo{person}{Nicholas~D Lane}, {and} \bibinfo{person}{Thomas Ploetz}.} \bibinfo{year}{2020}\natexlab{}.
\newblock \showarticletitle{Imutube: Automatic extraction of virtual on-body accelerometry from video for human activity recognition}.
\newblock \bibinfo{journal}{\emph{Proceedings of the ACM on Interactive, Mobile, Wearable and Ubiquitous Technologies}} \bibinfo{volume}{4}, \bibinfo{number}{3} (\bibinfo{year}{2020}), \bibinfo{pages}{1--29}.
\newblock


\bibitem[Landau and Lifshitz(1976)]%
        {dynacc}
\bibfield{author}{\bibinfo{person}{Lev~Davidovich Landau} {and} \bibinfo{person}{Evgenii~Mikhailovich Lifshitz}.} \bibinfo{year}{1976}\natexlab{}.
\newblock \bibinfo{booktitle}{\emph{Mechanics}}. Vol.~\bibinfo{volume}{1 (3rd ed.). Course of Theoretical Physics}.
\newblock \bibinfo{publisher}{Butterworth–Heinemann}. 126--130 pages.
\newblock


\bibitem[Lee and Joo(2024)]%
        {lee2024mocap}
\bibfield{author}{\bibinfo{person}{Jiye Lee} {and} \bibinfo{person}{Hanbyul Joo}.} \bibinfo{year}{2024}\natexlab{}.
\newblock \showarticletitle{Mocap Everyone Everywhere: Lightweight Motion Capture With Smartwatches and a Head-Mounted Camera}.
\newblock \bibinfo{journal}{\emph{arXiv preprint arXiv:2401.00847}} (\bibinfo{year}{2024}).
\newblock


\bibitem[Lee et~al\mbox{.}(2023)]%
        {questenvsim}
\bibfield{author}{\bibinfo{person}{Sunmin Lee}, \bibinfo{person}{Sebastian Starke}, \bibinfo{person}{Yuting Ye}, \bibinfo{person}{Jungdam Won}, {and} \bibinfo{person}{Alexander Winkler}.} \bibinfo{year}{2023}\natexlab{}.
\newblock \showarticletitle{QuestEnvSim: Environment-Aware Simulated Motion Tracking from Sparse Sensors}.
\newblock \bibinfo{journal}{\emph{arXiv preprint arXiv:2306.05666}} (\bibinfo{year}{2023}).
\newblock


\bibitem[Liang et~al\mbox{.}(2023)]%
        {hybridcap}
\bibfield{author}{\bibinfo{person}{Han Liang}, \bibinfo{person}{Yannan He}, \bibinfo{person}{Chengfeng Zhao}, \bibinfo{person}{Mutian Li}, \bibinfo{person}{Jingya Wang}, \bibinfo{person}{Jingyi Yu}, {and} \bibinfo{person}{Lan Xu}.} \bibinfo{year}{2023}\natexlab{}.
\newblock \showarticletitle{Hybridcap: Inertia-aid monocular capture of challenging human motions}. In \bibinfo{booktitle}{\emph{Proceedings of the AAAI Conference on Artificial Intelligence}}, Vol.~\bibinfo{volume}{37}. \bibinfo{pages}{1539--1548}.
\newblock


\bibitem[Loper et~al\mbox{.}(2015)]%
        {SMPL}
\bibfield{author}{\bibinfo{person}{Matthew Loper}, \bibinfo{person}{Naureen Mahmood}, \bibinfo{person}{Javier Romero}, \bibinfo{person}{Gerard Pons-Moll}, {and} \bibinfo{person}{Michael~J. Black}.} \bibinfo{year}{2015}\natexlab{}.
\newblock \showarticletitle{{SMPL}: A Skinned Multi-Person Linear Model}.
\newblock \bibinfo{journal}{\emph{ACM Trans. Graphics (Proc. SIGGRAPH Asia)}}  \bibinfo{volume}{34} (\bibinfo{date}{oct} \bibinfo{year}{2015}).
\newblock


\bibitem[Mahmood et~al\mbox{.}(2019)]%
        {AMASS}
\bibfield{author}{\bibinfo{person}{Naureen Mahmood}, \bibinfo{person}{Nima Ghorbani}, \bibinfo{person}{Nikolaus~F. Troje}, \bibinfo{person}{Gerard Pons-Moll}, {and} \bibinfo{person}{Michael~J. Black}.} \bibinfo{year}{2019}\natexlab{}.
\newblock \showarticletitle{AMASS: Archive of Motion Capture as Surface Shapes}. In \bibinfo{booktitle}{\emph{The IEEE International Conference on Computer Vision (ICCV)}}.
\newblock


\bibitem[Mollyn et~al\mbox{.}(2023)]%
        {imuposer}
\bibfield{author}{\bibinfo{person}{Vimal Mollyn}, \bibinfo{person}{Riku Arakawa}, \bibinfo{person}{Mayank Goel}, \bibinfo{person}{Chris Harrison}, {and} \bibinfo{person}{Karan Ahuja}.} \bibinfo{year}{2023}\natexlab{}.
\newblock \showarticletitle{IMUPoser: Full-Body Pose Estimation using IMUs in Phones, Watches, and Earbuds}. In \bibinfo{booktitle}{\emph{Proceedings of the 2023 CHI Conference on Human Factors in Computing Systems}}. \bibinfo{pages}{1--12}.
\newblock


\bibitem[Mur-Artal and Tard{\'o}s(2017)]%
        {orbslam2}
\bibfield{author}{\bibinfo{person}{Raul Mur-Artal} {and} \bibinfo{person}{Juan~D Tard{\'o}s}.} \bibinfo{year}{2017}\natexlab{}.
\newblock \showarticletitle{Orb-slam2: An open-source slam system for monocular, stereo, and rgb-d cameras}.
\newblock \bibinfo{journal}{\emph{IEEE transactions on robotics}} \bibinfo{volume}{33}, \bibinfo{number}{5} (\bibinfo{year}{2017}), \bibinfo{pages}{1255--1262}.
\newblock


\bibitem[Noitom({[n.\,d.]})]%
        {Noitom}
\bibfield{author}{\bibinfo{person}{Noitom}.} \bibinfo{year}{[n.\,d.]}\natexlab{}.
\newblock \bibinfo{title}{Perception Neuron series}.
\newblock \bibinfo{howpublished}{Website}.
\newblock
\newblock
\shownote{\url{https://www.noitom.com/}}.


\bibitem[Paige and Saunders(1982)]%
        {lsqr}
\bibfield{author}{\bibinfo{person}{Christopher~C. Paige} {and} \bibinfo{person}{Michael~A. Saunders}.} \bibinfo{year}{1982}\natexlab{}.
\newblock \showarticletitle{LSQR: An Algorithm for Sparse Linear Equations and Sparse Least Squares}.
\newblock \bibinfo{journal}{\emph{ACM Trans. Math. Softw.}} \bibinfo{volume}{8}, \bibinfo{number}{1} (\bibinfo{date}{mar} \bibinfo{year}{1982}), \bibinfo{pages}{43–71}.
\newblock
\showISSN{0098-3500}
\urldef\tempurl%
\url{https://doi.org/10.1145/355984.355989}
\showDOI{\tempurl}


\bibitem[Pan et~al\mbox{.}(2023)]%
        {robustcap}
\bibfield{author}{\bibinfo{person}{Shaohua Pan}, \bibinfo{person}{Qi Ma}, \bibinfo{person}{Xinyu Yi}, \bibinfo{person}{Weifeng Hu}, \bibinfo{person}{Xiong Wang}, \bibinfo{person}{Xingkang Zhou}, \bibinfo{person}{Jijunnan Li}, {and} \bibinfo{person}{Feng Xu}.} \bibinfo{year}{2023}\natexlab{}.
\newblock \showarticletitle{Fusing Monocular Images and Sparse IMU Signals for Real-time Human Motion Capture}. In \bibinfo{booktitle}{\emph{SIGGRAPH Asia 2023 Conference Papers}}. \bibinfo{pages}{1--11}.
\newblock


\bibitem[Pons-Moll et~al\mbox{.}(2010)]%
        {Pons2010}
\bibfield{author}{\bibinfo{person}{Gerard Pons-Moll}, \bibinfo{person}{Andreas Baak}, \bibinfo{person}{Thomas Helten}, \bibinfo{person}{Meinard Müller}, \bibinfo{person}{Hans-Peter Seidel}, {and} \bibinfo{person}{Bodo Rosenhahn}.} \bibinfo{year}{2010}\natexlab{}.
\newblock \showarticletitle{Multisensor-fusion for 3D full-body human motion capture}. In \bibinfo{booktitle}{\emph{2010 IEEE Computer Society Conference on Computer Vision and Pattern Recognition}}.
\newblock


\bibitem[Ponton et~al\mbox{.}(2023)]%
        {sparseposer}
\bibfield{author}{\bibinfo{person}{Jose~Luis Ponton}, \bibinfo{person}{Haoran Yun}, \bibinfo{person}{Andreas Aristidou}, \bibinfo{person}{Carlos Andujar}, {and} \bibinfo{person}{Nuria Pelechano}.} \bibinfo{year}{2023}\natexlab{}.
\newblock \showarticletitle{SparsePoser: Real-time Full-body Motion Reconstruction from Sparse Data}.
\newblock \bibinfo{journal}{\emph{ACM Transactions on Graphics}} \bibinfo{volume}{43}, \bibinfo{number}{1} (\bibinfo{year}{2023}), \bibinfo{pages}{1--14}.
\newblock


\bibitem[Pytorch({[n.\,d.]})]%
        {Pytorch}
\bibfield{author}{\bibinfo{person}{Pytorch}.} \bibinfo{year}{[n.\,d.]}\natexlab{}.
\newblock \bibinfo{title}{Pytorch}.
\newblock \bibinfo{howpublished}{Website}.
\newblock
\newblock
\shownote{\url{https://pytorch.org/}}.


\bibitem[Qin et~al\mbox{.}(2018)]%
        {vinsmono}
\bibfield{author}{\bibinfo{person}{Tong Qin}, \bibinfo{person}{Peiliang Li}, {and} \bibinfo{person}{Shaojie Shen}.} \bibinfo{year}{2018}\natexlab{}.
\newblock \showarticletitle{Vins-mono: A robust and versatile monocular visual-inertial state estimator}.
\newblock \bibinfo{journal}{\emph{IEEE Transactions on Robotics}} \bibinfo{volume}{34}, \bibinfo{number}{4} (\bibinfo{year}{2018}), \bibinfo{pages}{1004--1020}.
\newblock


\bibitem[Rey et~al\mbox{.}(2019)]%
        {rey2019let}
\bibfield{author}{\bibinfo{person}{Vitor~Fortes Rey}, \bibinfo{person}{Peter Hevesi}, \bibinfo{person}{Onorina Kovalenko}, {and} \bibinfo{person}{Paul Lukowicz}.} \bibinfo{year}{2019}\natexlab{}.
\newblock \showarticletitle{Let there be IMU data: generating training data for wearable, motion sensor based activity recognition from monocular RGB videos}. In \bibinfo{booktitle}{\emph{Adjunct proceedings of the 2019 ACM international joint conference on pervasive and ubiquitous computing and proceedings of the 2019 ACM international symposium on wearable computers}}. \bibinfo{pages}{699--708}.
\newblock


\bibitem[Shin et~al\mbox{.}(2023)]%
        {shin2023utilizing}
\bibfield{author}{\bibinfo{person}{Myungjin Shin}, \bibinfo{person}{Dohae Lee}, {and} \bibinfo{person}{In-Kwon Lee}.} \bibinfo{year}{2023}\natexlab{}.
\newblock \showarticletitle{Utilizing Task-Generic Motion Prior to Recover Full-Body Motion from Very Sparse Signals}.
\newblock \bibinfo{journal}{\emph{arXiv preprint arXiv:2308.15839}} (\bibinfo{year}{2023}).
\newblock


\bibitem[Skog et~al\mbox{.}(2010)]%
        {zupt}
\bibfield{author}{\bibinfo{person}{Isaac Skog}, \bibinfo{person}{Peter Handel}, \bibinfo{person}{John-Olof Nilsson}, {and} \bibinfo{person}{Jouni Rantakokko}.} \bibinfo{year}{2010}\natexlab{}.
\newblock \showarticletitle{Zero-velocity detection—An algorithm evaluation}.
\newblock \bibinfo{journal}{\emph{IEEE transactions on biomedical engineering}} \bibinfo{volume}{57}, \bibinfo{number}{11} (\bibinfo{year}{2010}), \bibinfo{pages}{2657--2666}.
\newblock


\bibitem[Sola(2017)]%
        {ESKF}
\bibfield{author}{\bibinfo{person}{Joan Sola}.} \bibinfo{year}{2017}\natexlab{}.
\newblock \showarticletitle{Quaternion kinematics for the error-state Kalman filter}.
\newblock \bibinfo{journal}{\emph{arXiv preprint arXiv:1711.02508}} (\bibinfo{year}{2017}).
\newblock


\bibitem[Takeda et~al\mbox{.}(2018)]%
        {takeda2018multi}
\bibfield{author}{\bibinfo{person}{Shingo Takeda}, \bibinfo{person}{Tsuyoshi Okita}, \bibinfo{person}{Paula Lago}, {and} \bibinfo{person}{Sozo Inoue}.} \bibinfo{year}{2018}\natexlab{}.
\newblock \showarticletitle{A multi-sensor setting activity recognition simulation tool}. In \bibinfo{booktitle}{\emph{Proceedings of the 2018 ACM International Joint Conference and 2018 International Symposium on Pervasive and Ubiquitous Computing and Wearable Computers}}. \bibinfo{pages}{1444--1448}.
\newblock


\bibitem[Trumble et~al\mbox{.}(2017)]%
        {TotalCapture}
\bibfield{author}{\bibinfo{person}{Matthew Trumble}, \bibinfo{person}{Andrew Gilbert}, \bibinfo{person}{Charles Malleson}, \bibinfo{person}{Adrian Hilton}, {and} \bibinfo{person}{John Collomosse}.} \bibinfo{year}{2017}\natexlab{}.
\newblock \showarticletitle{Total Capture: 3D Human Pose Estimation Fusing Video and Inertial Sensors}. In \bibinfo{booktitle}{\emph{2017 British Machine Vision Conference (BMVC)}}.
\newblock


\bibitem[Vaswani et~al\mbox{.}(2017)]%
        {vaswani2017attention}
\bibfield{author}{\bibinfo{person}{Ashish Vaswani}, \bibinfo{person}{Noam Shazeer}, \bibinfo{person}{Niki Parmar}, \bibinfo{person}{Jakob Uszkoreit}, \bibinfo{person}{Llion Jones}, \bibinfo{person}{Aidan~N Gomez}, \bibinfo{person}{{\L}ukasz Kaiser}, {and} \bibinfo{person}{Illia Polosukhin}.} \bibinfo{year}{2017}\natexlab{}.
\newblock \showarticletitle{Attention is all you need}.
\newblock \bibinfo{journal}{\emph{Advances in neural information processing systems}}  \bibinfo{volume}{30} (\bibinfo{year}{2017}).
\newblock


\bibitem[Vicon({[n.\,d.]})]%
        {Vicon}
\bibfield{author}{\bibinfo{person}{Vicon}.} \bibinfo{year}{[n.\,d.]}\natexlab{}.
\newblock \bibinfo{title}{Award Winning Motion Capture Systems}.
\newblock \bibinfo{howpublished}{Website}.
\newblock
\newblock
\shownote{\url{https://www.vicon.com/}}.


\bibitem[{von Marcard} et~al\mbox{.}(2017)]%
        {SIP}
\bibfield{author}{\bibinfo{person}{Timo {von Marcard}}, \bibinfo{person}{Bodo Rosenhahn}, \bibinfo{person}{Michael Black}, {and} \bibinfo{person}{Gerard Pons-Moll}.} \bibinfo{year}{2017}\natexlab{}.
\newblock \showarticletitle{Sparse Inertial Poser: Automatic 3D Human Pose Estimation from Sparse IMUs}.
\newblock \bibinfo{journal}{\emph{Computer Graphics Forum 36(2), Proceedings of the 38th Annual Conference of the European Association for Computer Graphics (Eurographics)}} (\bibinfo{year}{2017}).
\newblock


\bibitem[Winkler et~al\mbox{.}(2022)]%
        {questsim}
\bibfield{author}{\bibinfo{person}{Alexander Winkler}, \bibinfo{person}{Jungdam Won}, {and} \bibinfo{person}{Yuting Ye}.} \bibinfo{year}{2022}\natexlab{}.
\newblock \showarticletitle{QuestSim: Human Motion Tracking from Sparse Sensors with Simulated Avatars}. In \bibinfo{booktitle}{\emph{SIGGRAPH Asia 2022 Conference Papers}}. \bibinfo{pages}{1--8}.
\newblock


\bibitem[Xiao et~al\mbox{.}(2021)]%
        {xiao2021deep}
\bibfield{author}{\bibinfo{person}{Fanyi Xiao}, \bibinfo{person}{Ling Pei}, \bibinfo{person}{Lei Chu}, \bibinfo{person}{Danping Zou}, \bibinfo{person}{Wenxian Yu}, \bibinfo{person}{Yifan Zhu}, {and} \bibinfo{person}{Tao Li}.} \bibinfo{year}{2021}\natexlab{}.
\newblock \showarticletitle{A deep learning method for complex human activity recognition using virtual wearable sensors}. In \bibinfo{booktitle}{\emph{Spatial Data and Intelligence: First International Conference, SpatialDI 2020, Virtual Event, May 8--9, 2020, Proceedings 1}}. Springer, \bibinfo{pages}{261--270}.
\newblock


\bibitem[Xsens({[n.\,d.]})]%
        {Xsens}
\bibfield{author}{\bibinfo{person}{Xsens}.} \bibinfo{year}{[n.\,d.]}\natexlab{}.
\newblock \bibinfo{title}{Xsens 3D motion tracking}.
\newblock \bibinfo{howpublished}{Website}.
\newblock
\newblock
\shownote{\url{https://www.xsens.com/}}.


\bibitem[Yang et~al\mbox{.}(2021)]%
        {lobstr}
\bibfield{author}{\bibinfo{person}{Dongseok Yang}, \bibinfo{person}{Doyeon Kim}, {and} \bibinfo{person}{Sung-Hee Lee}.} \bibinfo{year}{2021}\natexlab{}.
\newblock \showarticletitle{Lobstr: Real-time lower-body pose prediction from sparse upper-body tracking signals}. In \bibinfo{booktitle}{\emph{Computer Graphics Forum}}, Vol.~\bibinfo{volume}{40}. Wiley Online Library, \bibinfo{pages}{265--275}.
\newblock


\bibitem[Ye et~al\mbox{.}(2022)]%
        {neural3points}
\bibfield{author}{\bibinfo{person}{Yongjing Ye}, \bibinfo{person}{Libin Liu}, \bibinfo{person}{Lei Hu}, {and} \bibinfo{person}{Shihong Xia}.} \bibinfo{year}{2022}\natexlab{}.
\newblock \showarticletitle{Neural3Points: Learning to Generate Physically Realistic Full-body Motion for Virtual Reality Users}. In \bibinfo{booktitle}{\emph{Computer Graphics Forum}}, Vol.~\bibinfo{volume}{41}. Wiley Online Library, \bibinfo{pages}{183--194}.
\newblock


\bibitem[Yi et~al\mbox{.}(2023)]%
        {EgoLocate}
\bibfield{author}{\bibinfo{person}{Xinyu Yi}, \bibinfo{person}{Yuxiao Zhou}, \bibinfo{person}{Marc Habermann}, \bibinfo{person}{Vladislav Golyanik}, \bibinfo{person}{Shaohua Pan}, \bibinfo{person}{Christian Theobalt}, {and} \bibinfo{person}{Feng Xu}.} \bibinfo{year}{2023}\natexlab{}.
\newblock \showarticletitle{EgoLocate: Real-time Motion Capture, Localization, and Mapping with Sparse Body-mounted Sensors}.
\newblock \bibinfo{journal}{\emph{ACM Transactions on Graphics (TOG)}} \bibinfo{volume}{42}, \bibinfo{number}{4}, Article \bibinfo{articleno}{76} (\bibinfo{year}{2023}), \bibinfo{numpages}{17}~pages.
\newblock


\bibitem[Yi et~al\mbox{.}(2022)]%
        {PIP}
\bibfield{author}{\bibinfo{person}{Xinyu Yi}, \bibinfo{person}{Yuxiao Zhou}, \bibinfo{person}{Marc Habermann}, \bibinfo{person}{Soshi Shimada}, \bibinfo{person}{Vladislav Golyanik}, \bibinfo{person}{Christian Theobalt}, {and} \bibinfo{person}{Feng Xu}.} \bibinfo{year}{2022}\natexlab{}.
\newblock \showarticletitle{Physical Inertial Poser (PIP): Physics-aware Real-time Human Motion Tracking from Sparse Inertial Sensors}. In \bibinfo{booktitle}{\emph{IEEE/CVF Conference on Computer Vision and Pattern Recognition (CVPR)}}.
\newblock


\bibitem[Yi et~al\mbox{.}(2021)]%
        {TransPose}
\bibfield{author}{\bibinfo{person}{Xinyu Yi}, \bibinfo{person}{Yuxiao Zhou}, {and} \bibinfo{person}{Feng Xu}.} \bibinfo{year}{2021}\natexlab{}.
\newblock \showarticletitle{TransPose: Real-time 3D Human Translation and Pose Estimation with Six Inertial Sensors}.
\newblock \bibinfo{journal}{\emph{ACM Transactions on Graphics}}  \bibinfo{volume}{40} (\bibinfo{date}{08} \bibinfo{year}{2021}).
\newblock


\bibitem[Young et~al\mbox{.}(2011)]%
        {imusim}
\bibfield{author}{\bibinfo{person}{Alexander~D Young}, \bibinfo{person}{Martin~J Ling}, {and} \bibinfo{person}{Damal~K Arvind}.} \bibinfo{year}{2011}\natexlab{}.
\newblock \showarticletitle{IMUSim: A simulation environment for inertial sensing algorithm design and evaluation}. In \bibinfo{booktitle}{\emph{Proceedings of the 10th ACM/IEEE International Conference on Information Processing in Sensor Networks}}. IEEE, \bibinfo{pages}{199--210}.
\newblock


\bibitem[Zheng et~al\mbox{.}(2023)]%
        {Zheng_2023_ICCV}
\bibfield{author}{\bibinfo{person}{Xiaozheng Zheng}, \bibinfo{person}{Zhuo Su}, \bibinfo{person}{Chao Wen}, \bibinfo{person}{Zhou Xue}, {and} \bibinfo{person}{Xiaojie Jin}.} \bibinfo{year}{2023}\natexlab{}.
\newblock \showarticletitle{Realistic Full-Body Tracking from Sparse Observations via Joint-Level Modeling}. In \bibinfo{booktitle}{\emph{Proceedings of the IEEE/CVF International Conference on Computer Vision (ICCV)}}. \bibinfo{pages}{14678--14688}.
\newblock


\end{thebibliography}

\appendix
\clearpage
\section{Derivation of Fictitious Acceleration}
In this section, we derive the mathematical formula for the fictitious acceleration of a single leaf joint within the human body's root frame during motion.
Following the notation and subscript conventions established in the main paper, the position of the leaf joint in the root frame can be calculated with respect to the world frame as:
\begin{equation}\label{eq:p}
    \boldsymbol{p}_{RL}=\boldsymbol{R}^{-1}_{WR}(\boldsymbol{p}_{WL}-\boldsymbol{p}_{WR}),
\end{equation}
where $W$, $L$, and $R$ stand for the world, leaf, and root frame respectively.
By taking the time derivative of both sides of Eq.~\ref{eq:p}, we obtain:
\begin{equation}\label{eq:pdot}
\begin{aligned}
    \dot{\boldsymbol{p}}_{RL}
    &=\dot{(\boldsymbol{R}^{-1}_{WR})}(\boldsymbol{p}_{WL}-\boldsymbol{p}_{WR}) + \boldsymbol{R}^{-1}_{WR}(\dot{\boldsymbol{p}}_{WL} - \dot{\boldsymbol{p}}_{WR})\\
    &=\dot{\boldsymbol{R}}_{WR}^T(\boldsymbol{p}_{WL}-\boldsymbol{p}_{WR}) + \boldsymbol{R}^{-1}_{WR}(\dot{\boldsymbol{p}}_{WL} - \dot{\boldsymbol{p}}_{WR})\\
    &=(\boldsymbol{R}_{WR}[\boldsymbol{\omega}_{RR}]_\times)^T(\boldsymbol{p}_{WL}-\boldsymbol{p}_{WR}) + \boldsymbol{R}^{-1}_{WR}(\dot{\boldsymbol{p}}_{WL} - \dot{\boldsymbol{p}}_{WR})\\
    &=-[\boldsymbol{\omega}_{RR}]_\times\left(\boldsymbol{R}^{-1}_{WR}(\boldsymbol{p}_{WL}-\boldsymbol{p}_{WR})\right)+ \boldsymbol{R}^{-1}_{WR}(\dot{\boldsymbol{p}}_{WL} - \dot{\boldsymbol{p}}_{WR})\\
    &=-[\boldsymbol{\omega}_{RR}]_\times\boldsymbol{p}_{RL}+ \boldsymbol{R}^{-1}_{WR}(\dot{\boldsymbol{p}}_{WL} - \dot{\boldsymbol{p}}_{WR}).
\end{aligned}
\end{equation}
In the derivation above, it is important to note that we utilize the orthogonality property of a rotation matrix expressed as $\boldsymbol{R}^{-1}=\boldsymbol{R}^T$, as well as the relationship for the time derivative $\dot{\boldsymbol{R}}=\boldsymbol{R}[\boldsymbol\omega]_\times$, where $\boldsymbol\omega$ represents the local body-frame angular velocity. Furthermore, we take into account the antisymmetry property of a skew-symmetric matrix, written as $[\boldsymbol\omega]_\times^T=-[\boldsymbol\omega]_\times$.
Continuing with our analysis, we further compute the time derivative of both sides of Eq.~\ref{eq:pdot}, resulting in:
\begin{scriptsize}  
\begin{equation}\label{eq:pddot}
\begin{aligned}
    \ddot{\boldsymbol{p}}_{RL}
    &=-[\dot{\boldsymbol{\omega}}_{RR}]_\times\boldsymbol{p}_{RL}-[\boldsymbol{\omega}_{RR}]_\times\dot{\boldsymbol{p}}_{RL} + \dot{(\boldsymbol{R}^{-1}_{WR})}(\dot{\boldsymbol{p}}_{WL} - \dot{\boldsymbol{p}}_{WR})+ \boldsymbol{R}^{-1}_{WR}(\ddot{\boldsymbol{p}}_{WL}-\ddot{\boldsymbol{p}}_{WR})\\
    &=-[\dot{\boldsymbol{\omega}}_{RR}]_\times\boldsymbol{p}_{RL}-[\boldsymbol{\omega}_{RR}]_\times\dot{\boldsymbol{p}}_{RL}+ (\boldsymbol{R}_{WR}[\boldsymbol{\omega}_{RR}]_\times)^T(\dot{\boldsymbol{p}}_{WL}-\dot{\boldsymbol{p}}_{WR})-\boldsymbol{a}_{RR}+\boldsymbol{a}_{RL}\\
    &=-[\dot{\boldsymbol{\omega}}_{RR}]_\times\boldsymbol{p}_{RL}-[\boldsymbol{\omega}_{RR}]_\times\dot{\boldsymbol{p}}_{RL} - [\boldsymbol{\omega}_{RR}]_\times(\dot{\boldsymbol{p}}_{RL}+[\boldsymbol{\omega}_{RR}]_\times\boldsymbol{p}_{RL})-\boldsymbol{a}_{RR}+\boldsymbol{a}_{RL}\\
    &=-[\dot{\boldsymbol{\omega}}_{RR}]_\times\boldsymbol{p}_{RL}-2[\boldsymbol{\omega}_{RR}]_\times\dot{\boldsymbol{p}}_{RL} - [\boldsymbol{\omega}_{RR}]_\times^2\boldsymbol{p}_{RL}-\boldsymbol{a}_{RR}+\boldsymbol{a}_{RL}.
\end{aligned}
\end{equation}
\end{scriptsize}
Note that in the third line, we need to substitute Eq~\ref{eq:pdot} into the third term.
Finally, the result obtained in Eq.~\ref{eq:pddot} is essentially identical to Eq. 1 presented in the main paper, indicating that the fictitious acceleration consists of four terms, following the same order as in Eq.~\ref{eq:pddot}, which are named as the Euler, Coriolis, centrifugal, and linear terms.
For further reading, see~\cite{dynacc}.

\section{More experiments on IMU Synthesis}
In the proposed IMU synthesis technique, we implement an IMU fusion algorithm based on the Error-State Kalman Filter (ESKF) as outlined in~\cite{ESKF}. In this section, we present additional experiments conducted on the IMU synthesis technique (Sec.~\ref{sec:app1}) and the implementation of ESKF-based IMU fusion (Sec.~\ref{sec:app2}).
\subsection{Evaluation on IMU Synthesis}\label{sec:app1}
The effectiveness of our IMU synthesis method is demonstrated by the final improvement observed in the trained models, as depicted in Tab. 2 and Fig. 7b in the main paper.
Readers may find interest in understanding how our synthesized IMU noise exhibits similarities to those in real IMU measurements. 
However, directly comparing two noise sequences can be challenging. 
In Fig. 8 of the main paper, we visually illustrate how closely our noise patterns align with real IMU measurements.
In this section, we provide more quantitative evaluation results.
\par
As directly computing the mean-square error (MSE) between two noise sequences may not be the most suitable approach, we endeavor to assess the \textit{cosine similarity} between the synthesized acceleration and the real IMU measurement in the frequency domain.
Tab.~\ref{tab:synnoise} shows the similarity in magnitudes of low- and high-frequency noise between synthetic and real sensor accelerations. 
Our method has a higher similarity score, indicating that it generally yields more realistic measurements, especially in the high-frequency component, which is consistent with the acceleration data visualization in Fig. 8.

\begin{table}[]
\caption{Cosine similarity between the synthetic and real IMU accelerations in the frequency domain. Different IMU synthesizing methods are tested using the TotalCapture dataset. We analyze both low- and high-frequency components with a cut-off frequency set at 10Hz. A score closer to 1 indicates a higher level of similarity between the synthetic and real data.}
\label{tab:synnoise}
\resizebox{0.9\linewidth}{!}{
\begin{tabular}{cccc}
\toprule
Test band              & DIP    & TransPose/PIP   & Ours            \\ \midrule
Low frequency (<10Hz)  & 0.9117 & 0.9092          & \textbf{0.9124} \\
High frequency (>10Hz) & 0.7156 & 0.6832          & \textbf{0.7758} \\
Full band              & 0.7866 & 0.8334          & \textbf{0.8588} \\
\bottomrule
\end{tabular}}
\end{table}

\begin{figure}
    \centering
    \includegraphics[width=\linewidth]{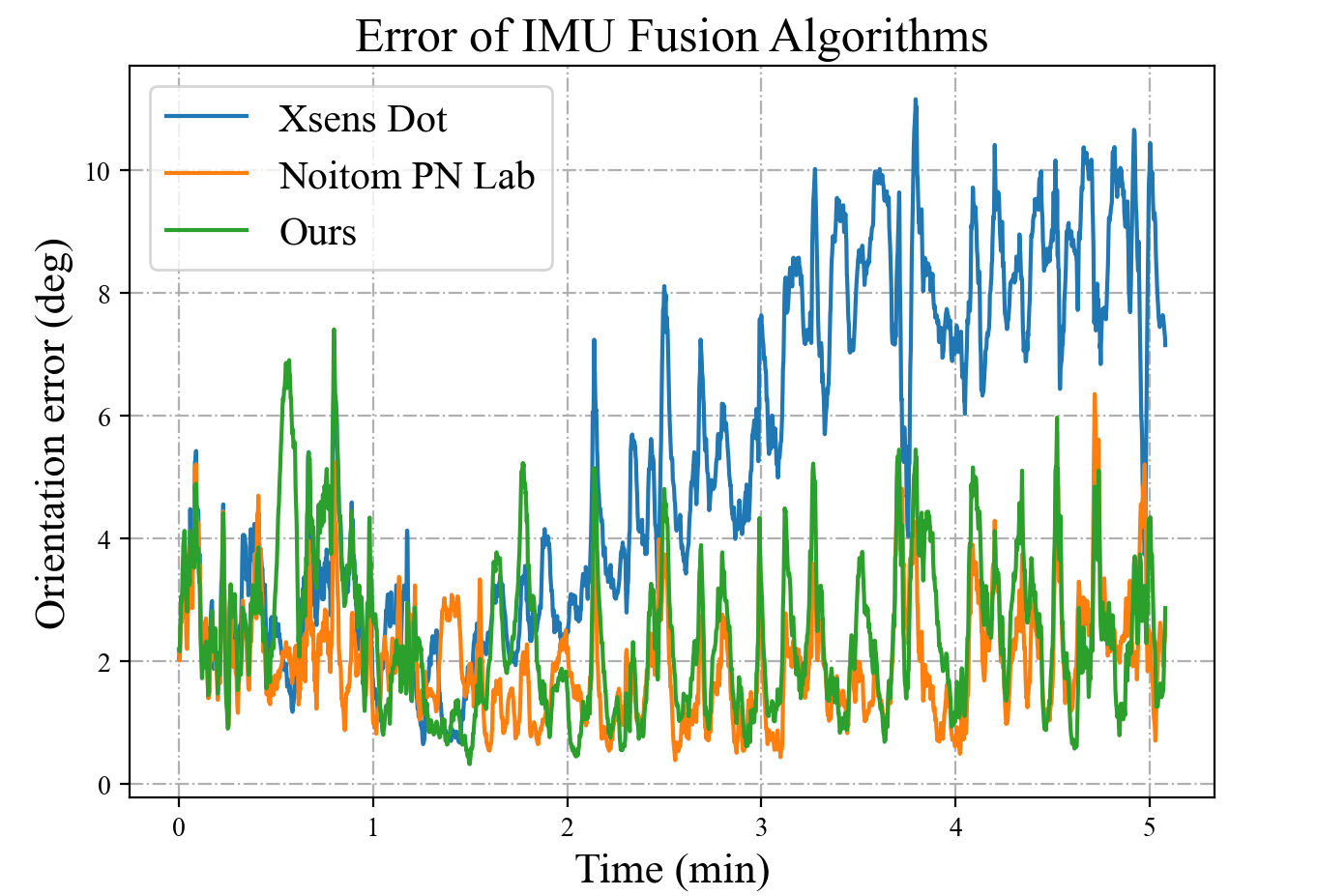}
    \caption{Orientation error of different IMU fusion algorithms in our recording. Mean error (degrees): Xsens Dot 5.3; Noitom PN Lab 2.0; Ours 2.4.}
    \label{fig:imufusion}
\end{figure}

\subsection{Evaluation on IMU Fusion}\label{sec:app2}
To evaluate our implementation of the ESKF-based IMU fusion algorithm~\cite{ESKF}, we conduct comparisons with Xsens~\cite{Xsens} and Noitom~\cite{Noitom} black-box orientation filter integrated within their sensors (we use Xsens Dot IMU and Noitom PN Lab IMU).
We record 5-minute IMU measurements with ground-truth orientations obtained from an optical system. 
We then compare the orientation outputs of the Xsens, Noitom, and our algorithm with the ground truth. 
The result is shown in Fig.~\ref{fig:imufusion}.
Our algorithm achieves a comparable accuracy compared to Noitom PN Lab, surpassing Xsens Dot. 
We have observed that Xsens Dot experiences drift along the gravity axis, possibly due to their utilization of a different algorithm for interpreting magnetic measurements.

\section{Discussion on Failure Cases}
In the supplementary video, we present three failure cases of the proposed approach. In this section, we provide an in-depth discussion on these failure cases.
\paragraph{Motion with insensible acceleration} Our technique excels at capturing acceleration-dependent motion by establishing a better understanding of the correlation between human motion and acceleration measurements. However, when the subject performs a very slow motion where the acceleration is almost indistinguishable from the sensor noise, our method cannot accurately estimate the motion. This is evident in the first failure case shown in the video, where the subject intentionally throws a punch very slowly. In such cases, the measurements of acceleration and orientation almost remain constant during the motion, where our algorithm fails to accurately capture the motion. However, as the speed of the motion increases, our method is able to capture the motion accurately.
\paragraph{Motion on uneven terrain} Our method employs a physical optimizer adopted from PIP~\cite{PIP}, which assumes that the floor is flat. Consequently, when the person steps on a stair, the reconstructed motion briefly shows the person walking onto a higher height. However, the physical optimization quickly pulls the person back to the floor due to gravity. To accurately capture motion on uneven terrain, the physical optimizer requires a scene scan as input to account for the unflattened ground.
\paragraph{Motion under severely distorted magnetic environment} The third failure case in the video shows that our method does not perform optimally when capturing human motion in an environment with a severely distorted magnetic field. This is because the inertial sensors we use assume a constant magnetic field to filter the yaw orientation. When the human enters an area with a different magnetic north direction, the sensor's orientation output becomes inaccurate, which in turn affects our results. It is important to note that this does not mean the environment must be entirely free of magnetic distortion. As long as the magnetic field in the environment is spatially constant (even if it differs from the earth's geomagnetic field), our system can still perform well.

\end{document}